\documentstyle[12pt,amsfonts]{article}

\def\hybrid{\topmargin 0pt      \oddsidemargin 0pt
        \headheight 0pt \headsep 0pt
        \textwidth 6.25in       
        \textheight 9.5in       
        \marginparwidth .875in
        \parskip 5pt plus 1pt   \jot = 1.5ex}
\makeatletter
\def\marginnote#1{}
\newcount\hour
\newcount\minute
\newtoks\amorpm
\hour=\time\divide\hour by60
\minute=\time{\multiply\hour by60 \global\advance\minute
by-\hour}\edef\standardtime{{\ifnum\hour<12
\global\amorpm={am}%
        \else\global\amorpm={pm}\advance\hour by-12 \fi
        \ifnum\hour=0 \hour=12 \fi
        \number\hour:\ifnum\minute<10
0\fi\number\minute\the\amorpm}}
\edef\militarytime{\number\hour:\ifnum\minute<10
0\fi\number\minute}
 
\def\draftlabel#1{{\@bsphack\if@filesw {\let\thepage\relax
   \xdef\@gtempa{\write\@auxout{\string
      \newlabel{#1}{{\@currentlabel}{\thepage}}}}}\@gtempa
   \if@nobreak \ifvmode\nobreak\fi\fi\fi\@esphack}
        \gdef\@eqnlabel{#1}}
\def\@eqnlabel{}
\def\@vacuum{}
\def\draftmarginnote#1{\marginpar{\raggedright\scriptsize\tt#1}}
\def\draft{\oddsidemargin -.5truein
        \def\@oddfoot{\sl preliminary draft \hfil
        \rm\thepage\hfil\sl\today\quad\militarytime}
        \let\@evenfoot\@oddfoot \overfullrule 3pt
        \let\label=\draftlabel
        \let\marginnote=\draftmarginnote
 
\def\@eqnnum{(\theequation)\rlap{\kern\marginparsep\tt\@eqnlabel}%
\global\let\@eqnlabel\@vacuum}  }
 
\def\numberbysection{\@addtoreset{equation}{section}
        \def\theequation{\thesection.\arabic{equation}}}

\def\underline#1{\relax\ifmmode\@@underline#1\else
 $\@@underline{\hbox{#1}}$\relax\fi}

\def\eqnarray{\stepcounter{equation}\let\@currentlabel=\theequation
\global\@eqnswtrue
\global\@eqcnt\z@\tabskip\@centering\let\\=\@eqncr
$$\halign to \displaywidth\bgroup\@eqnsel\hskip\@centering
  $\displaystyle\tabskip\z@{##}$&\global\@eqcnt\@ne 
  \hfil$\displaystyle{\hbox{}##\hbox{}}$\hfil
  &\global\@eqcnt\tw@ $\displaystyle\tabskip\z@
  {##}$\hfil\tabskip\@centering&\llap{##}\tabskip\z@\cr}
\def\lefteqn#1{\hbox to 2em{$\displaystyle #1$\hss}}
\def\ad{\mathop{\rm ad}\nolimits}
\def\id{\mathop{\rm id}\nolimits}
\def\r#1{\raisebox{-0.3ex}{$\displaystyle
  \mathop{\rho}^{\scriptscriptstyle (#1)}$}{}}
\def\t#1{\raisebox{-0.1ex}{$\displaystyle
  \mathop{\tau}^{\scriptscriptstyle (#1)}$}{}}
\def\s#1{\raisebox{-0.1ex}{$\displaystyle
  \mathop{\sigma}^{\scriptscriptstyle (#1)}$}{}}
\mathchardef\by="202
\def\mbar#1{\kern 0.1em\overline{\kern -0.1em #1 \kern -0.1em} 
  \kern 0.1em}
\makeatother
\numberbysection
\hybrid

\begin{document}

\begin{titlepage}
\nopagebreak
\begin{flushright}
LPTENS-96/69\\
hep-th/9612081
\\
December 1996
\end{flushright}
\vskip 5cm
\begin{center}
{\large\bf
MAXIMALLY NONABELIAN TODA SYSTEMS}
\vglue 1  true cm
{\bf Alexander V. Razumov}\\
{\footnotesize Institute for High Energy Physics,
142284, Protvino, Moscow region, Russia\footnote{E-mail:
razumov@mx.ihep.su}}\\  
and\\
{\bf Mikhail V. Saveliev}\footnote{
On leave of absence from the Institute for High Energy Physics,
142284, Protvino, Moscow region, Russia; e-mail: saveliev@mx.ihep.su}
\\
{\footnotesize Laboratoire de Physique Th\'eorique de
l'\'Ecole Normale Sup\'erieure\footnote{Unit\'e Propre du
Centre National de la Recherche Scientifique,
associ\'ee \`a l'\'Ecole Normale Sup\'erieure et \`a l'Universit\'e
de Paris-Sud},\\
24 rue Lhomond, 75231 Paris C\'EDEX 05, ~France\footnote{
E-mail: saveliev@physique.ens.fr}}

\medskip
\end{center}

\vfill
\begin{abstract}
\baselineskip .4 true cm
\noindent
A detailed consideration of the maximally nonabelian Toda systems
based on the classical semisimple Lie groups is given. The explicit
expressions for the general solution of the corresponding equations
are obtained.
\end{abstract}
\vfill
\end{titlepage}
\baselineskip .5 true cm

\section{Introduction}

Last two decades the Toda systems permanently attract great attention
of the physicists and mathematicians due to their integrability and
deep links with a number of problems in the theory of differential
equations, differential and algebraic geometry, Lie algebras, Lie
groups and their representations, etc.; and relevance to many
problems of modern theoretical and mathematical physics.  In
particular, they arise in a natural way in various approaches of the
particle physics (e.g., field theory models and supergravity
including black holes and $p$-branes business) and statistical
mechanics (e.g., correlations in the inhomogeneous $XY$-model at
infinite temperature). Whereas there is a lot of papers devoted to
classical and quantum behaviour of abelian Toda systems, nonabelian
Toda systems remain in many aspects unexplored. It is mainly caused
by the absence of nontrivial examples, for which one can write the
general solution for the equations under consideration in a rather
explicit form. However, there exists a class of nonabelian Toda
systems having a very simple structure. We call these systems
maximally nonabelian Toda systems by the following reason.

A Toda system is related to some Lie group $G$ whose Lie algebra
${\frak g}$ is equipped with a ${\Bbb Z}$-gradation, and hence there
is defined the subgroup $\tilde H$ corresponding to the
subalgebra formed by the zero grade elements. The Toda fields
parametrise a mapping from ${\Bbb R}^2$ or ${\Bbb C}$ to $\tilde
H$. If the subgroup $\tilde H$ is abelian, we deal with an
abelian Toda system, otherwise we have a nonabelian Toda system. In
the case of the trivial ${\Bbb Z}$-gradation the subgroup $\tilde
H$ coincides with $G$ and we come to the Wess-Zumino-Novikov-Witten
equations.  The maximally nonabelian Toda systems correspond to the
case when the subgroup $\tilde H$ does not coincide with $G$ and
is not a proper subgroup of any subgroup of $G$ generated by the zero
grade subspace for some nontrivial ${\Bbb Z}$-gradation of ${\frak
g}$.

In the present paper we give a detailed consideration of the
maximally nonabelian Toda systems associated with the classical
semisimple finite dimensional Lie groups.

\section{Toda systems and their integration}

\subsection{${\Bbb Z}$-gradations}

The starting  point for the construction
\cite{LSa92,RSa94,RSa96,RSa96a} of Toda equations is a complex Lie
group whose Lie algebra is endowed with a ${\Bbb Z}$-gradation. In
the first two sections we give some necessary information about
${\Bbb Z}$-gradations of the complex semisimple Lie algebras, for
more details see, for example, \cite{GOV94,RSa96}.  Recall that a Lie
algebra ${\frak g}$ is said to be endowed with a ${\Bbb Z}$-gradation
if there is given a representation of ${\frak g}$ as a direct sum
\[
{\frak g} = \bigoplus_{a \in {\Bbb Z}} {\frak g}_a,
\]
where $[{\frak g}_a, {\frak g}_b] \subset {\frak g}_{a+b}$ for all $a,
b \in {\Bbb Z}$.

Let $G$ be a complex Lie group, and ${\frak g}$ be its Lie
algebra. For a given ${\Bbb Z}$-gradation of ${\frak g}$ introduce
the following subalgebras of ${\frak g}$:
\[
\tilde{\frak h} \equiv {\frak g}_0, \qquad
\tilde{\frak n}_- \equiv \bigoplus_{a < 0} {\frak g}_a, \qquad
\tilde{\frak n}_+ \equiv \bigoplus_{a > 0} {\frak g}_a.
\]
Here and henceforth we use tildes to have the notations different
from those usually used in the case of the so called canonical
gradation of complex semisimple Lie algebras.

Denote by $\tilde H$ and $\tilde N_\pm$ the connected Lie
subgroups of $G$ corresponding to the subalgebras $\tilde{\frak
h}$ and $\tilde{\frak n}_\pm$ respectively.  Suppose that
$\tilde H$ and $\tilde N_\pm$ are closed subgroups of $G$
and, moreover,
\begin{eqnarray*}
&\tilde H \cap \tilde N_\pm = \{e\}, \qquad \tilde N_-
\cap \tilde N_+ = \{e\}, \\
&\tilde N_- \cap \tilde H \tilde N_+ = \{e\}, \qquad
\tilde N_- \tilde H \cap \tilde N_+ = \{e\}.
\end{eqnarray*}
where $e$ is the unit element of $G$. This is true, in particular,
for the finite dimensional reductive Lie groups, see, for example,
\cite{Hum75}. The set $\tilde N_-  \tilde H \tilde N_+$
is an open subset of $G$. Suppose that
$ G = \mbar{\tilde N_- \tilde H \tilde N_+}$,
where the bar means the topological closure. This is again true for
the finite dimensional reductive Lie groups.  Thus, in the case under
consideration for any element $a$ which belongs to the dense set
$\tilde N_- \tilde H \tilde N_+$ one can write the
following unique decomposition
\begin{equation}
a = n_- h n_+^{-1}, \label{2.1}
\end{equation}
where $n_- \in \tilde N_-$, $h \in \tilde H$ and $n_+ \in
\tilde N_+$. Decomposition (\ref{2.1}) is called the Gauss
decomposition. Note that the Gauss decomposition (\ref{2.1}) is the
principal tool used in the group-algebraic integration procedure for
Toda equations.

There is a simple classification of possible ${\Bbb Z}$-gradations
for complex semi\-simple Lie algebras. In this case for any ${\Bbb
Z}$-gradation of a such an algebra ${\frak g}$, there exists a
unique element $q \in {\frak g}$ which has the following property. An
element $x \in {\frak g}$ belongs to the subspace ${\frak g}_a$ if
and only if $[q, x] = ax$.  This can be written as
\[
{\frak g}_a = \{x \in {\frak g} \mid [q, x] = ax \}.
\]
The element $q$ is called the grading operator. It is clearly
semisimple, and
\[
\exp(2\pi \sqrt{-1} \ad q) = \id_{\frak g}.
\]
On the other hand, any semisimple element of ${\frak g}$ which
possesses this property defines a ${\Bbb Z}$-gradation of ${\frak
g}$.

Since the grading operator $q$ is semisimple, one can always point out
a Cartan subalgebra of ${\frak g}$ which contains $q$ \cite{Dix74}.
Let ${\frak h}$ be such a Cartan subalgebra, and $\Delta$ be the root
system of ${\frak g}$ with respect to ${\frak h}$. For any element
$x$ of the root subspace ${\frak g}^\alpha$ corresponding to the
root $\alpha \in \Delta$, one has $[q, x] = \langle \alpha, q \rangle x$,
where $\langle \alpha, q \rangle$ means the action of the element
$\alpha \in {\frak g}^*$ on the element $q \in {\frak g}$.
Hence, for any root $\alpha \in \Delta$ the number $\langle \alpha, q
\rangle$ is an integer. Furthermore, if we choose a base $ \Pi =
\{\alpha_1, \ldots, \alpha_r\}$ of $\Delta$ corresponding to the Weyl
chamber whose closure contains $q$, then the integers $s_i \equiv
\langle \alpha_i, q \rangle$ are nonnegative. The grading subspace ${\frak
g}_a$, $a \ne 0$, is the direct sum of the root subspaces ${\frak
g}_\alpha$ corresponding to the roots $\alpha = \sum_{i=1}^r n_i
\alpha_i$ with $\sum_{i=1}^r n_i s_i = a$. The subspace ${\frak g}_0$,
besides of the root subspaces corresponding to the roots $\alpha =
\sum_{i=1}^r n_i \alpha_i$ with $\sum_{i=1}^r n_i s_i = 0$, includes
the Cartan subalgebra ${\frak h}$.

It can be shown that, up to a possible reordering related to the
freedom in the renumbering of the elements of $\Pi$, the numbers
$s_i$ do not depend neither on the choice of a Cartan subalgebra
containing $q$, nor on the choice of a base possessing the above
described property. In other words, a ${\Bbb Z}$-gradation of a
semisimple Lie algebra ${\frak g}$ is described by nonnegative
integer labels associated with the vertices of the corresponding
Dynkin diagram. Two ${\Bbb Z}$-gradations are connected by an inner
automorphism of ${\frak g}$ if and only if they have the same set of
labels.  Two ${\Bbb Z}$-gradations are connected by an `external'
automorphism of ${\frak g}$ if and only if the corresponding sets of
labels are connected by an automorphism of the Dynkin diagram.

Let now ${\frak h}$ be a Cartan subalgebra of a semisimple Lie
algebra ${\frak g}$, $\Delta$ be the root system of ${\frak g}$ with
respect to ${\frak h}$, and $\Pi = \{\alpha_1, \ldots, \alpha_r\}$ be
a base of $\Delta$. Denote by $h_i$ and $x_{\pm i}$, $i = 1, \ldots,
r$, the corresponding Cartan and Chevalley generators of ${\frak g}$.
For any set of $r$ nonnegative numbers $s_i$ the element
\begin{equation}
q = \sum_{i,j=1}^r (k^{-1})_{ij} s_j h_i, \label{2.5}
\end{equation}
where $k = (k_{ij})$ is the Cartan matrix of ${\frak g}$, is the
grading operator of some ${\Bbb Z}$-gradation of ${\frak g}$. Here
the numbers $s_i$ are the corresponding labels of the Dynkin diagram.
The canonical gradation of ${\frak g}$ arises when one chooses all the
number $s_i$ equal to 1. Thus, there is a bijective correspondence
between the sets of nonnegative integer labels of the Dynkin diagram
of a semisimple Lie algebra and the classes of its conjugated ${\Bbb
Z}$-gradations. 

If all the labels of the Dynkin diagram of a semisimple Lie algebra
${\frak g}$ are different from zero, then the subgroup ${\frak g}_0$
coincides with some Cartan subalgebra of ${\frak g}$.  In this case
the subgroup $\tilde H$ is abelian and we deal with the so called
abelian Toda equations. In all other cases the subgroup $\tilde
H$ is nonabelian and the corresponding Toda equations are called
nonabelian. In particular, if all the labels are equal to zero, then
there is only one grading subspace ${\frak g}_0 = {\frak g}$. In this
case the corresponding Toda equations coincide with the
Wess-Zumino-Novikov-Witten equations. In the present paper
we consider the case when only one of the labels is different from zero;
the corresponding equations are called here maximally nonabelian
Toda equations.

Sometimes it is more convenient to consider, instead of a semisimple
Lie algebra, some reductive Lie algebra which contains this semisimple
Lie algebra. Here we use ${\Bbb Z}$-gradations
defined by the following procedure. Recall that a reductive Lie
algebra ${\frak g}$ can be represented as the direct product of the
center $Z({\frak g})$ of ${\frak g}$ and a semisimple subalgebra
${\frak g}'$  of ${\frak g}$. Choose some ${\Bbb Z}$-gradation of
${\frak g}'$ and denote the corresponding grading operator by $q$. It
is clear that $q$ is the grading operator of some ${\Bbb
Z}$-gradation of the whole Lie algebra ${\frak g}$. Moreover, 
the subspaces ${\frak g}_a$, $a \ne 0$, coincide with the subspaces
${\frak g}'_a$, and the subspace ${\frak g}_0$ is the direct sum of
the subspace ${\frak g}'_0$ and the center $Z({\frak g})$.

\subsection{{\rm ${\frak sl}$(2, ${\Bbb C}$)}-subalgebras}

It is often useful to consider ${\Bbb Z}$-gradations of a Lie algebra
${\frak g}$ associated with embeddings of the Lie algebra ${\frak
sl}(2, {\Bbb C})$ into ${\frak g}$. Recall that the Lie algebra
${\frak sl}(2, {\Bbb C})$ is a complex simple Lie algebra formed by
all traceless $2 \by 2$ matrices. This Lie algebra is of rank 1, and
the Cartan and Chevalley generators satisfy the following commutation
relations
\begin{eqnarray}
&[x_+, x_-] = h, \label{2.9} \\
&[h, x_-] = -2 x_-, \qquad [h, x_+] = 2 x_+. \label{2.10}
\end{eqnarray}
By an embedding of ${\frak sl}(2, {\Bbb C})$ into ${\frak g}$ we mean
a nontrivial homeomorphism from ${\frak sl}(2, {\Bbb C})$ into ${\frak
g}$. The images of the elements $h$ and $x_\pm$ under the
homomorphism, defining the embedding under consideration, are denoted
usually by the same letters. The image of the whole Lie algebra
${\frak sl}(2, \Bbb C)$ is called an ${\frak sl}(2, {\Bbb
C})$-subalgebra of ${\frak g}$. For a given embedding of ${\frak
sl}(2, {\Bbb C})$ into ${\frak g}$, the adjoint representation of
${\frak g}$ defines the representation of the Lie algebra ${\frak
sl}(2, {\Bbb C})$ in ${\frak g}$. From the properties of the finite
dimensional representations of ${\frak sl}(2, {\Bbb C})$ it follows
the element $h$ of ${\frak g}$ must be semisimple, and the elements
$x_\pm$ must be nilpotent. Moreover, it is clear that $\exp (2 \pi
\sqrt{-1} \ad h) = \id_{\frak g}$. 
Therefore, the element $h$ can be used as the grading operator
defining some ${\Bbb Z}$-gradation of ${\frak g}$. It can be shown
that in the case when ${\frak g}$ is semisimple, the labels of the
corresponding Dynkin diagram can be equal only to 0, 1, and 2. In
particular, if the labels are equal only to 0 or 2, we deal with the
so called integral embedding. In this case it is natural to consider
the ${\Bbb Z}$-gradation defined by the grading operator $q = h/2$.

\subsection{Toda equations}

Let $M$ be either a real two dimensional manifold, or a complex one
dimensional manifold. Choose some local coordinates $z^-$ and $z^+$
on $M$. In the complex case we assume that $z^+ = \mbar{z^-}$.
Denote the partial derivatives over $z^+$ and $z^-$ by $\partial_+$
and $\partial_-$ respectively. Consider a ${\Bbb Z}$-graded complex
semisimple Lie algebra ${\frak g}$. Let $l$ be a positive integer, such
that the grading subspaces ${\frak g}_a$ for $-l < a < 0$ and $0 <
a < l$ are trivial, and $c_-$ and $c_+$ be some fixed elements of
the subspaces ${\frak g}_{-l}$ and ${\frak g}_{+l}$ respectively.
Restrict ourselves to the case when $G$ is a matrix Lie group. In
this case the Toda equations are the matrix partial differential
equations of the form 
\begin{equation}
\partial_+(\gamma^{-1} \partial_- \gamma) = [c_-, \gamma^{-1} c_+
\gamma], \label{2.2}
\end{equation}
where $\gamma$ is a mapping from $M$ to $\tilde H$.

Let $h_\pm$ be some elements of $\tilde H$, and the mapping
$\gamma$ satisfies the Toda equations. It is easy to get convinced
that the mapping 
\[
\gamma' = h_+^{-1} \gamma h_-
\]
satisfies the Toda equations (\ref{2.2}) with the elements $c_\pm$
replaced by the elements
\begin{equation}
c'_\pm = h_\pm^{-1} c_\pm h_\pm. \label{2.11}
\end{equation}
Io this sense, the Toda equations determined by the elements $c_\pm$
and $c'_\pm$ which are connected by the above relation, are
equivalent. 

Denote by $\tilde H_\pm$ the subgroups of $\tilde H$ defined by
\begin{equation}
\tilde H_\pm = \{ h \in \tilde H \mid h c_\pm h^{-1} = c_\pm
\}. \label{2.12}
\end{equation}
The Toda equations are invariant with respect to the transformations
\begin{equation}
\gamma' = \xi_+^{-1} \gamma \xi_- \label{2.13}
\end{equation}
where $\xi_\pm$ are arbitrary mappings from $M$ to the subgroups
$\tilde H_\pm$, satisfying the relations $\partial_\mp \xi_\pm =
0$.

\subsection{Integration scheme} \label{int}

To obtain the general solution of Toda equations one can use the
following procedure \cite{LSa92,RSa94,RSa96,RSa96a}. Choose some
mappings $\gamma_\pm$ from $M$ to $\tilde H$ such that
\begin{equation}
\partial_\mp \gamma_\pm = 0. \label{2.6}
\end{equation}
Integrate the equations
\begin{equation}
\mu^{-1}_\pm \partial_\pm \mu_\pm = \gamma_\pm x_\pm \gamma_\pm^{-1},
\qquad \partial_\mp \mu_\pm = 0. \label{2.3}
\end{equation}
The solutions of the above equations are fixed by the conditions
\begin{equation}
\mu_\pm(p) = a_\pm, \label{2.4}
\end{equation}
where $p$ is some fixed point of $M$, and $a_\pm$ are some
elements of $G$. The mappings $\mu_\pm$ satisfying equations
(\ref{2.3}) and conditions (\ref{2.4}) are unique and take values in
the sets $a_\pm \tilde N_\pm$. The Gauss decomposition
(\ref{2.1}) induces the corresponding decomposition of mappings from
$M$ to $G$. In particular, one obtains
\begin{equation}
\mu^{-1}_+ \mu_- = \nu_- \eta \nu_+^{-1}, \label{2.7}
\end{equation}
where the mapping $\eta$ takes values in $\tilde H$, and the
mappings $\nu_\pm$ take values in $\tilde N_\pm$.
It can be shown that the mapping
\begin{equation}
\gamma = \gamma^{-1}_+ \eta \gamma_- \label{2.8}
\end{equation}
satisfies the Toda equations, and any solution to this equation can be
obtained by the described procedure. Note that almost all solutions
of the Toda equations can be obtained using the mappings $\mu_\pm$
submitted to relation (\ref{2.4}) with $a_\pm \in \tilde N_\pm$,
or, in other words, using the mappings $\mu_\pm$ taking values in the
subgroups $\tilde N_\pm$ \cite{RSa96,RSa96a}. 

\section{Complex general linear group}

\subsection{Equations}

We begin the consideration of maximally nonabelian Toda systems with
the case of the Lie group ${\rm GL}(r+1, {\Bbb C})$. Recall that the
corresponding Lie algebra ${\frak gl}(r+1, {\Bbb C})$ is reductive
and can be represented as the direct product of the simple Lie
algebra ${\frak sl}(r+1, {\Bbb C})$ and a one dimensional Lie algebra
isomorphic to ${\frak gl}(1, {\Bbb C})$ and composed by the $(r+1)
\by (r+1)$ complex matrices which are multiplies of the unit matrix.

The Lie algebra ${\frak sl}(r+1, {\Bbb C})$ is of type $A_r$.
Let $d$ be a fixed integer such that $1 \le d \le r$. Consider the
${\Bbb Z}$-gradation of ${\frak sl}(r+1, {\Bbb C})$ arising when we
choose the labels of the corresponding Dynkin diagram equal to
zero except the label $s_d$ which is chosen equal to 1. Construct the
grading operator associated with this gradation using
relation (\ref{2.5}) and the well known explicit expression for the inverse 
of the Cartan matrix, see e.g., \cite{GGr78,RSa96}.
{}From relation (\ref{2.5}) it follows that the grading operator in
the case under consideration has the form
\[
q = \frac{1}{r+1} \left[(r+1-d) \sum_{i=1}^{d-1} i h_i + d
\sum_{i=d}^r (r+1-i) h_i \right].
\]
It is convenient to take as a Cartan subalgebra for ${\frak sl}(r+1,
{\Bbb C})$ the subalgebra consisting of diagonal $(r+1) \by (r+1)$
matrices with zero trace. Here the standard choice of the Cartan
generators is 
\[
h_i = e_{i,i} - e_{i+1, i+1},
\]
where the matrices $e_{i, j}$ are defined by
\begin{equation}
(e_{i,j})_{kl} \equiv \delta_{ik} \delta_{jl}. \label{3.34}
\end{equation}
With such a choice of Cartan generators we obtain
\[
q = \frac{1}{r+1} \left[(r+1-d) \sum_{i=1}^d e_{i,i} - d
\sum_{i=d+1}^{r+1} e_{i,i} \right].
\]
Thus, the grading operator has the following block matrix form:
\begin{equation}
q = \left( \begin{array}{cc}
\frac{m_2}{m_1 + m_2} I_{m_1} & 0 \\
0  & \frac{-m_1}{m_1 + m_2} I_{m_2}
\end{array} \right), \label{3.1}
\end{equation}
where $m_1 = d$ and $m_2 = r+1-d$, so that $m_1 + m_2 = r+1$.  Here
and henceforth $I_m$ denotes the unit $m \by m$ matrix.  We will use
this grading operator to define a ${\Bbb Z}$-gradation of the Lie
algebra ${\frak gl}(r+1, {\Bbb C})$.  It is not difficult to show
that in this case we have three grading subspaces, ${\frak g}_0$ and
${\frak g}_{\pm 1}$. To describe these subspaces, it is convenient to
consider $(r+1) \by (r+1)$ matrices as $2 \by 2$ block matrices $x =
(x_{ij})$, where $x_{ij}$ are $m_i \by m_j$ matrices.  Then the
subspace ${\frak g}_0 = \tilde {\frak h}$ consists of all block
diagonal matrices, and the subspaces ${\frak g}_{-1} =
\tilde {\frak n}_-$ and ${\frak g}_{+1} = \tilde {\frak n}_+$
are formed by all block strictly lower and upper triangular matrices
respectively. In other words, one can say that the subspace ${\frak
g}_a$ consists of the matrices $x = (x_{ij})$ where only the
blocks $x_{ij}$ with $j-i = a$ are different from zero.

It is easy to describe the corresponding subgroups of ${\rm
GL}(r+1, {\Bbb C})$.  The subgroup $\tilde H$ is formed by all
block diagonal nondegenerate matrices, and the subgroups $\tilde
N_-$ and $\tilde N_+$ consist respectively of all block upper and
lower triangular matrices with unit matrices on the diagonal.

Proceed now to the consideration of the corresponding Toda equations.
The general form of the elements $c_\pm$ is
\begin{equation}
c_- = \left( \begin{array}{cc}
0 & 0 \\
C_- & 0
\end{array} \right), \qquad 
c_+ = \left( \begin{array}{cc}
0 & C_+ \\
0 & 0
\end{array} \right). \label{3.35}
\end{equation}
Parametrise the mapping $\gamma$ as
\[
\gamma = \left( \begin{array}{cc}
\beta_1 & 0 \\
0 & \beta_2
\end{array} \right),
\]
where the mappings $\beta_1$ and $\beta_2$ take values in the groups
${\rm GL}(m_1, {\Bbb C})$ and ${\rm GL}(m_2, {\Bbb C})$
respectively. Using now the relation
\[
\gamma^{-1} c_+ \gamma = \left( \begin{array}{cc}
0 & \beta^{-1}_1 C_+ \beta_2 \\
0 & 0
\end{array} \right),
\]
we can write the Toda equations in the form
\begin{eqnarray}
&&\partial_+ (\beta_1^{-1} \partial_- \beta_1) = -\beta_1^{-1} C_+
\beta_2 C_-, \label{3.2} \\
&&\partial_+ (\beta_2^{-1} \partial_- \beta_2) = C_- \beta_1^{-1} C_+
\beta_2. \label{3.3}
\end{eqnarray}
Recall that the Toda equations defined by the elements $c_\pm$ and
$c'_\pm$ connected by relation (\ref{2.11}) are equivalent.  For the
case under consideration this statement is equivalent to saying that
equations (\ref{3.2}), (\ref{3.3}) are determined by fixing the ranks
of the matrices $C_-$ and $C_+$.

Let us try to associate the ${\Bbb Z}$-gradation of ${\frak gl}(r+1,
{\Bbb C})$ which is considered in this section with some embedding of
the Lie algebra ${\frak sl}(2, {\Bbb C})$ into ${\frak gl}(r+1, {\Bbb
C})$.  It is clear that the Cartan generator $h$ of the corresponding
${\frak sl}(2, {\Bbb C})$-subalgebra should coincide with $2q$. The
Chevalley generators $x_-$ and $x_+$ should belong to the subspaces
${\frak g}_{-1}$ and ${\frak g}_{+1}$ respectively.  Therefore, we
can write for $x_-$ and $x_+$ the block matrix representation
\begin{equation}
x_- = \left( \begin{array}{cc}
0 & 0 \\
X_- & 0
\end{array} \right), \qquad 
x_+ = \left( \begin{array}{cc}
0 & X_+ \\
0 & 0
\end{array} \right). \label{3.36}
\end{equation}
To satisfy relation (\ref{2.9}) we should have
\[
X_+ X_- = \frac{2 m_2}{m_1+m_2} I_{m_1}, \qquad X_- X_+ = \frac{2
m_1}{m_1 + m_2} I_{m_2}. 
\]
{}From this equalities it follows that the rank $r(X_+ X_-)$ is equal
to $m_1$ and the rank $r(X_- X_+)$ is equal to $m_2$. On the other
hand, 
\[
r(X_+ X_-) \le {\rm min}(r(X_+), r(X_-)) \le {\rm min}(m_1,
m_2). 
\]
Hence, one has $m_1 \le {\rm min}(m_1, m_2)$ and $m_2 \le {\rm
min}(m_1, m_2)$. This is possible only if $m_1 = m_2 = m$, and in
this case $r(X_+) = m$ and $r(X_-) = m$. It can be easily shown that
the equality $m_1 = m_2$ is a sufficient condition for the existence
of an ${\frak sl}(2, {\Bbb C})$-subalgebra in question.

If the condition $m_1 = m_2$ is satisfied, then without any loss of
generality, we can choose 
\begin{equation}
h = \left( \begin{array}{cc}
I_m & 0 \\
0 & -I_m
\end{array} \right), \qquad
x_- = \left( \begin{array}{cc}
0 & 0 \\
I_m & 0
\end{array} \right), \qquad
x_+ = \left( \begin{array}{cc}
0 & I_m \\
0 & 0
\end{array} \right). \label{3.9}
\end{equation}
Now, with $c_\pm = x_\pm$, one comes to the Toda equations of the
form
\begin{eqnarray}
&&\partial_+ (\beta_1^{-1} \partial_- \beta_1) = -\beta_1^{-1}
\beta_2, \label{3.7} \\ 
&&\partial_+ (\beta_2^{-1} \partial_- \beta_2) = \beta_1^{-1} \beta_2,
\label{3.8} 
\end{eqnarray}
where the mappings $\beta_1$ and $\beta_2$ take values in the Lie
group ${\rm GL}(m, {\Bbb C})$. In this case the subgroups $\tilde
H_\pm$ defined by (\ref{2.12}) are isomorphic to the Lie group ${\rm
GL}(m, {\Bbb C})$ and are composed of the block matrices of the form
\begin{equation}
h = \left( \begin{array}{cc}
h_{11} & 0 \\
0 & h_{11}
\end{array} \right). \label{3.37}
\end{equation}
Recall that these subgroups determine the symmetry transformations
(\ref{2.13}) of the Toda equations.

\subsection{General solution}

In accordance with the general scheme described in section 2, to
obtain the general solution for equations (\ref{3.2}), (\ref{3.3}) one
should start with the mappings $\gamma_\pm$ taking values in the
subgroup $\tilde H$ and satisfying relations (\ref{2.6}). Write
these mappings in the block matrix form 
\[
\gamma_\pm = \left( \begin{array}{cc}
\beta_{\pm 1} & 0 \\
0 & \beta_{\pm 2}
\end{array} \right)
\]
where the mappings $\beta_{\pm 1}$ take values in the group ${\rm
GL}(m_1, {\Bbb C})$, and the mappings $\beta_{\pm 2}$ take values in
the group ${\rm GL}(m_2, {\Bbb C})$. Now we have to integrate
equations (\ref{2.3}). Since almost all solutions can be obtained
using the mappings $\mu_\pm$ taking values in the subgroups
$\tilde N_\pm$, we choose for $\mu_\pm$ the parametrisation of
the form
\begin{equation}
\mu_- = \left( \begin{array}{cc}
I_{m_1} & 0 \\
\mu_{-21} & I_{m_2}
\end{array} \right), \qquad
\mu_+ = \left( \begin{array}{cc}
I_{m_1} & \mu_{+12} \\
0 & I_{m_2}
\end{array} \right), \label{3.6}
\end{equation}
with the mapping $\mu_{-21}$ taking values in the space of $m_2
\by m_1$ matrices and the mapping $\mu_{+12}$ taking values in the
space of $m_1 \by m_2$ matrices.  This representation allows to
reduce equations (\ref{2.3}) to the equations
\begin{eqnarray*}
&&\partial_- \mu_{-21} = \beta_{-2} C_- \beta_{-1}^{-1}, \qquad
\partial_+ \mu_{-21} = 0, \\
&&\partial_+ \mu_{+12} = \beta_{+1} C_+ \beta_{+2}^{-1}, \qquad
\partial_- \mu_{+12} = 0.
\end{eqnarray*}
The general solution to these equations is
\begin{eqnarray}
&&\mu_{-21} (z^-) = m_{-21} + \int_0^{z^-} dy^- \beta_{-2} (y^-) C_-
\beta_{-1}^{-1} (y^-), \label{3.4} \\ 
&&\mu_{+12} (z^+) = m_{+12} + \int_0^{z^+} dy^+ \beta_{+1} (y^+) C_+
\beta_{+2}^{-1} (y^+), \label{3.5}
\end{eqnarray}
where $m_{-21}$ and $m_{+12}$ are arbitrary $m_2 \by m_1$ and $m_1
\by m_2$ matrices respectively.

Consider now the Gauss decomposition (\ref{2.7}). From (\ref{3.6})
one gets
\[
\mu_+^{-1} \mu_- = \left( \begin{array}{cc}
I_{m_1} - \mu_{+12} \mu_{-21} & - \mu_{+12} \\
\mu_{-21} & I_{m_2}
\end{array} \right).
\]
Parametrising the mapping $\eta$ as 
\[
\eta = \left( \begin{array}{cc}
\eta_{11} & 0 \\
0 & \eta_{22}
\end{array} \right),
\]
and using (\ref{a.4}), (\ref{a.5}) we find that
\[
\eta_{11} = I_{m_1} - \mu_{+12} \mu_{-21}, \qquad \eta_{22} = I_{m_2} +
\mu_{-21} (I_{m_1} - \mu_{+12} \mu_{-21})^{-1} \mu_{+12}.
\]
Note that the mapping $\mu_+^{-1} \mu_-$ has the Gauss decomposition
(\ref{2.7}) only at those points of $M$ for which
\[
\det (I_{m_1} - \mu_{+12}(z^+) \mu_{-21}(z^-)) \ne 0.
\]
Now, using (\ref{2.8}), we come to the following expression for the
general solution of equations (\ref{3.2}), (\ref{3.3}):
\begin{eqnarray}
&&\beta_1 = \beta_{+1}^{-1} (I_{m_1} - \mu_{+12} \mu_{-21}) \beta_{-1},
\label{3.10} \\
&&\beta_2 = \beta_{+2}^{-1} \left[ I_{m_2} + \mu_{-21}
(I_{m_1} - \mu_{+12} \mu_{-21})^{-1} \mu_{+12} \right] \beta_{-2},
\label{3.11} 
\end{eqnarray}
where the mappings $\mu_{-21}$ and $\mu_{+12}$ are given by
(\ref{3.4}) and (\ref{3.5}).

Proceed now to the case when the considered ${\Bbb Z}$-gradation is
associated with the ${\frak sl}(2, {\Bbb C})$-subalgebra defined by
(\ref{3.9}). Recall that with the choice $c_\pm = x_\pm$ the Toda
equations have form (\ref{3.7}), (\ref{3.8}). Introduce the notation
$\beta = \beta_1$, and write expression (\ref{3.10}) as
\[
\beta(z^-, z^+) = \zeta_{+1}(z^+) \zeta_{-1}(z^-) + \zeta_{+2}(z^+)
\zeta_{-2} (z^-),
\]
where
\bgroup \def\arraystretch{1.4} \[
\begin{array}{lcl}
\zeta_{+1} = \beta_{+1}^{-1}, &\qquad& \zeta_{-1} = \beta_{-1}, \\
\zeta_{+2} = -\beta_{+1}^{-1} \mu_{+12}, &\qquad& \zeta_{-2} =
\mu_{-21} \beta_{-1}.
\end{array}
\] \egroup
{}From (\ref{3.7}) one obtains
\[
\beta_2 = - \partial_+ \partial_- \beta + \partial_+ \beta \beta^{-1}
\partial_- \beta.
\]
Consider the family of quasideterminants 
\begin{equation}
\Delta_{i} = \left| \begin{array}{rrrr}
\beta & \partial_- \beta & \cdots & \partial_-^{i-1} \beta \\
\partial_+ \beta & \partial_+ \partial_- \beta & \cdots &
\partial_+ \partial_-^{i-1} \beta \\
\vdots \hfil & \vdots \hfil & \ddots & \vdots \hfil \\
\partial_+^{i-1} \beta & \partial_+^{i-1} \partial_- \beta & \cdots &
\partial_+^{i-1} \partial_-^{i-1} \beta
\end{array} \right|_{ii}. \label{3.12}
\end{equation}
The definition of a quasideterminant which is used in our work is
given by (\ref{a.19}). A more general definition of quasideterminants
and investigation of their properties can be found in
\cite{GRe91,GRe92}. It follows from (\ref{a.18}) that
\[
\beta_1 = \Delta_{1}, \qquad \beta_2 = - \Delta_{2}.
\]
It is another form of writing the general solution to equations
(\ref{3.7}), (\ref{3.8}).

\subsection{Generalisation}

In this section we consider the Toda equations also based on the Lie
group ${\rm GL}(r+1, {\Bbb C})$ but with more general ${\Bbb
Z}$-gradations of ${\frak gl}(r+1, {\Bbb C})$. Strictly speaking, the
systems under consideration are not maximally nonabelian Toda
systems. Nevertheless, it is possible to find for their general
solution some interesting explicit expression. Moreover, maximally
nonabelian Toda systems based on classical Lie groups different from
${\rm GL}(r+1, {\Bbb C})$, can be considered as reductions of the
systems which are investigated here. 

Let $d_1$ and $d_2$ be two positive integers such that $1 \le d_1 <
d_2 \le r$. Consider the ${\Bbb Z}$-gradation of ${\frak gl}(r+1,
{\Bbb C})$ corresponding to the case when we choose all the labels
of the Dynkin diagram equal to zero, except the labels $s_{d_1}$
and $s_{d_2}$ which are chosen equal to 1. The corresponding grading
operator can be constructed as follows. Denote the grading operator
given by relation (\ref{3.1}) by $q_d$, then it is clear that the
grading operator $q$ in question is $q = q_{d_1} + q_{d_2}$. The
explicit form of $q$ is
\begin{equation}
q = \left( \begin{array}{ccc}
\frac{m_2 + 2 m_3}{m_1 + m_2 + m_3} I_{m_1} & 0 & 0 \\
0 & \frac{-m_1 + m_3}{m_1 + m_2 + m_3} I_{m_2} & 0 \\
0 & 0 &  \frac{- 2 m_1 - m_2}{m_1 + m_2 + m_3} I_{m_3}
\end{array} \right), \label{3.14}
\end{equation} 
where $m_1 = d_1$, $m_2 = d_2 - d_1$ and $m_3 = r + 1 - d_2$, so that
$m_1 + m_2 + m_3 = r + 1$. 

In this example we consider $(r+1) \by (r+1)$
matrices as $3 \by 3$ block matrices $x = (x_{ij})$ with $x_{ij}$
being $m_i \by m_j$ matrices. With respect to the ${\Bbb
Z}$-gradation of \hbox{${\frak gl}(r+1, {\Bbb C})$} defined by the grading
operator $q$ given by (\ref{3.14}), there are five nontrivial grading
subspaces ${\frak g}_{\pm 2}$, ${\frak g}_{\pm 1}$ and ${\frak g}_0$.
Here the subspace ${\frak g}_a$ is formed by the block matrices $x =
(x_{ij})$ where only the blocks $x_{ij}$ with $j - i = a$ are
different from zero.

The subalgebra $\tilde {\frak h}$ consists of all block diagonal
matrices, and the subspaces $\tilde {\frak n}_-$ and $\tilde
{\frak n}_+$ are formed by all block strictly lower and upper
triangular matrices respectively.  The subgroup $\tilde H$ is
formed by all block diagonal nondegenerate matrices, and the
subgroups $\tilde N_-$ and $\tilde N_+$ consist respectively
of all block upper and lower triangular matrices with unit matrices
on the diagonal.

In the case under consideration the general form of the elements
$c_\pm \in {\frak g}_{\pm 1}$ is
\[
c_- = \left( \begin{array}{ccc}
0 & 0 & 0\\
C_{-1} & 0 & 0\\
0 & C_{-2} & 0
\end{array} \right), \qquad 
c_+ = \left( \begin{array}{ccc}
0 & C_{+1} & 0\\
0 & 0 & C_{+2}\\
0 & 0 & 0 
\end{array} \right).
\]
Parametrising the mapping $\gamma$ as
\[
\gamma = \left( \begin{array}{ccc}
\beta_1 & 0 & 0\\
0 & \beta_2 & 0\\ 
0 & 0 & \beta_3
\end{array} \right),
\]
we come to the following Toda equations
\begin{eqnarray}
&&\partial_+(\beta^{-1}_1 \partial_- \beta_1) = - \beta^{-1}_1 C_{+1}
\beta_2 C_{-1}, \label{3.16} \\
&&\partial_+(\beta^{-1}_2 \partial_- \beta_2) = - \beta^{-1}_2 C_{+2}
\beta_3 C_{-2} + C_{-1} \beta_1^{-1} C_{+1} \beta_2, \label{3.17} \\
&&\partial_+(\beta^{-1}_3 \partial_- \beta_3) =  C_{-2} \beta^{-1}_2
C_{+2} \beta_3, \label{3.18}
\end{eqnarray}
where the mappings $\beta_i$, $i = 1,2,3$, take values in the Lie
groups ${\rm GL}(m_i, {\Bbb C})$.

Find now the conditions which guarantee the existence of an ${\frak
sl}(2, {\Bbb C})$-sub\-algeb\-ra of ${\frak gl}(r+1, {\Bbb C})$ giving
the ${\Bbb Z}$-gradation under consideration. Actually we are
interested only in an integral embedding of ${\frak sl}(2, {\Bbb
C})$.  Therefore, the Chevalley generators $x_-$ and $x_+$ should
belong to the subspaces ${\frak g}_{-1}$ and ${\frak g}_{+1}$
respectively, and the Cartan generator $h$ should coincide with $2q$.
In this case relations (\ref{2.10}) are satisfied.  Write for
$x_-$ and $x_+$ the block matrix representation
\[
x_- = \left( \begin{array}{ccc}
0 & 0 & 0\\
X_{-1} & 0 & 0\\
0 & X_{-2} & 0
\end{array} \right), \qquad 
x_+ = \left( \begin{array}{ccc}
0 & X_{+1} & 0\\
0 & 0 & X_{+2}\\
0 & 0 & 0 
\end{array} \right).
\]
To satisfy relation (\ref{2.9}) we should have
\begin{eqnarray}
&&X_{+1} X_{-1} = 2 \frac{m_2 + 2 m_3}{m_1 + m_2 + m_3} I_{m_1},
\label{3.19} \\
&&X_{+2} X_{-2} - X_{-1} X_{+1} = 2 \frac{-m_1 + m_3}{m_1 + m_2 + m_3}
I_{m_2}, \label{3.20} \\
&&X_{-2} X_{+2} = 2 \frac{2 m_1 + m_2}{m_1 + m_2 + m_3} I_{m_3}.
\label{3.21} 
\end{eqnarray}
{}From (\ref{3.19}) it follows that $m_1 \le m_2$ and that the ranks
$r(X_{\pm 1})$ are equal to $m_1$. On the other hand, relation
(\ref{3.21}) implies $m_3 \le m_2$ and $r(X_{\pm 2}) = m_3$.
Multiplying (\ref{3.20}) from the left by $X_{+1}$ and taking into
account (\ref{3.19}), one obtains
\[
X_{+1} X_{+2} X_{-2} = 2 \frac{-m_1 + m_2 + 3m_3}{m_1 + m_2 + m_3}
X_{+1}. 
\]
This relation gives that $m_1 \le m_3$.
Similarly, multiplying (\ref{3.20}) from the right by $X_{+2}$ and
taking into account (\ref{3.21}), we get the relation
\[
X_{-1} X_{+1} X_{+2} = 2 \frac{3 m_1 + m_2 - m_3}{m_1 + m_2 + m_2}
X_{+2},
\] 
which implies that $m_3 \le m_1$. Thus, one can satisfy relations
(\ref{3.19})--(\ref{3.21}) only if $m_1 = m_3$. It is easy to see
that the conditions $m_1 = m_3$ and $m_1 \le m_2$ are sufficient
conditions for the existence of an ${\frak sl}(2, {\Bbb
C})$-subalgebra leading to the ${\Bbb Z}$-gradation under
consideration.

In the simplest and most symmetric case arising when $m_1 = m_2 = m_3
= m$, we can take, without any loose of generality, as the sought for
${\frak sl}(2, {\Bbb C})$-subalgebra the subalgebra generated by the
elements
\begin{eqnarray}
&&h = \left( \begin{array}{ccc}
2 I_m & 0 & 0\\
0 & 0 & 0\\
0 & 0 & -2 I_m
\end{array} \right), \label{3.25} \\
&&x_- = \left( \begin{array}{ccc}
0 & 0 & 0\\
\sqrt{2} I_m & 0 & 0\\
0 & \sqrt{2} I_m & 0
\end{array} \right), \qquad
x_+ = \left( \begin{array}{ccc}
0 & \sqrt{2} I_m & 0\\
0 & 0 & \sqrt{2} I_m\\
0 & 0 & 0
\end{array} \right). \label{3.26}
\end{eqnarray}
With the choice $c_\pm = x_\pm/\sqrt{2}$ one comes to the Toda
equations of the form 
\begin{eqnarray}
&&\partial_+(\beta^{-1}_1 \partial_- \beta_1) = - \beta^{-1}_1
\beta_2, \label{3.22} \\ 
&&\partial_+(\beta^{-1}_2 \partial_- \beta_2) = - \beta^{-1}_2
\beta_3 + \beta_1^{-1} \beta_2, \label{3.23} \\
&&\partial_+(\beta^{-1}_3 \partial_- \beta_3) =  \beta^{-1}_2
\beta_3, \label{3.24}
\end{eqnarray}
where the mappings $\beta_i$, $i=1,2,3$, take values in the Lie
group ${\rm GL}(m, {\Bbb C})$.

Return now to the case of arbitrary $m_1$, $m_2$ and $m_3$.
To obtain the general solution of equations (\ref{3.16})--(\ref{3.18})
we start with the mappings $\gamma_{\pm}$ which are parametrised as
\[
\gamma_\pm = \left( \begin{array}{ccc}
\beta_{\pm 1} & 0 & 0\\
0 & \beta_{\pm 2} & 0\\
0 & 0 & \beta_{\pm 3}
\end{array} \right).
\]
Write for the mappings $\mu_\pm$ the following representation
\[
\mu_- = \left( \begin{array}{ccc}
I_{m_1} & 0 & 0 \\
\mu_{-21} & I_{m_2} & 0\\
\mu_{-31} & \mu_{-32} & I_{m_3}
\end{array} \right), \qquad
\mu_+ = \left( \begin{array}{ccc}
I_{m_1} & \mu_{+12} & \mu_{+13} \\
0 & I_{m_2} & \mu_{+23} \\
0 & 0 & I_{m_3}
\end{array} \right).
\]
Then equations (\ref{2.3}) take the form
\bgroup \def\arraystretch{1.4} \[
\begin{array}{lcl}
\partial_- \mu_{-21} = \beta_{-2} C_{-1} \beta_{-1}^{-1}, &\qquad&
\partial_+ \mu_{-21} = 0, \\
\partial_- \mu_{-32} = \beta_{-3} C_{-2} \beta_{-2}^{-1}, &\qquad&
\partial_+ \mu_{-32} = 0, \\
\partial_- \mu_{-31} = \mu_{-32} \beta_{-2} C_{-1}
\beta_{-1}^{-1}, &\qquad& \partial_+ \mu_{-31} = 0,\\
\partial_+ \mu_{+12} = \beta_{+1} C_{+1} \beta_{+2}^{-1}, &\qquad&
\partial_- \mu_{+12} = 0, \\
\partial_+ \mu_{+23} = \beta_{+2} C_{+2} \beta_{+3}^{-1}, &\qquad&
\partial_- \mu_{+23} = 0, \\
\partial_+ \mu_{+13} = \mu_{+12} \beta_{+2} C_{+2}
\beta_{+3}^{-1}, &\qquad& \partial_- \mu_{+13} = 0.
\end{array} 
\] \egroup
The general solution to these equations is
\begin{eqnarray*}
&&\mu_{-21} (z^-) = m_{-21} + \int_0^{z^-} dy_1^- \beta_{-2} (y_1^-)
C_{-1} \beta_{-1}^{-1} (y_1^-), \\
&&\mu_{-32} (z^-) = m_{-32} + \int_0^{z^-} dy_1^- \beta_{-3} (y_1^-)
C_{-2} \beta_{-2}^{-1} (y_1^-), \\
&&\mu_{-31} (z^-) = m_{-31} \\
&& \hskip 0.5em {} + \int_0^{z^-} dy^-_2 \left( m_{-32} +
\int_0^{y^-_2} dy^-_1 \beta_{-3} (y^-_1) C_{-2} \beta_{-2}^{-1}
(y^-_1) \right) \beta_{-2} (y^-_2) C_{-1} \beta_{-1}^{-1} (y^-_2),\\
&&\mu_{+12} (z^+) = m_{+12} + \int_0^{z^+} dy_1^+ \beta_{+1} (y_1^+)
C_{+1} \beta_{+2}^{-1} (y_1^+), \\
&&\mu_{+23} (z^+) = m_{+23} + \int_0^{z^+} dy_1^+ \beta_{+2} (y_1^+)
C_{+2} \beta_{+3}^{-1} (y_1^+), \\
&& \mu_{+13} (z^+) = m_{+13} \\
&& \hskip 0.5em{} + \int_0^{z^+} dy^+_2 \left( m_{+12} +
\int_0^{y^+_2} dy^+_1 \beta_{+1} (y^+_1) C_{+1} \beta_{+2}^{-1}
(y^+_1) \right) \beta_{+2} (y^+_2) C_{+2} \beta_{+3}^{-1} (y^+_2), 
\end{eqnarray*}
where $m_{-21}$, $m_{-32}$, $m_{-31}$, $m_{+12}$, $m_{+23}$ and
$m_{+13}$ are arbitrary constant matrices.

The next step of the integration procedure is to obtain from the
Gauss decomposition (\ref{2.7}) the mapping $\eta$. Using the relation
\[
\mu_+^{-1} = \left( \begin{array}{ccc}
I_{m_1} & - \mu_{+12} & - (\mu_{+13} - \mu_{+12} \mu_{+23}) \\
0 & I_{m_2} & -\mu_{+23} \\
0 & 0 & I_{m_3}
\end{array} \right),
\]
we get for the blocks determining the mapping $\mu_+^{-1} \mu_-$ the
following expressions 
\begin{eqnarray*}
&&(\mu_+^{-1} \mu_-)_{11} = I_{m_1} - \mu_{+12} \mu_{-21} - (\mu_{+13}
- \mu_{+12} \mu_{+23}) \mu_{-31}, \\
&&(\mu_+^{-1} \mu_-)_{12} =  - \mu_{+12} - (\mu_{+13} - \mu_{+12}
\mu_{+23}) \mu_{-32}, \\
&&(\mu_+^{-1} \mu_-)_{13} = - (\mu_{+13} - \mu_{+12} \mu_{+23}), \\ 
&&(\mu_+^{-1} \mu_-)_{21} = \mu_{-21} - \mu_{+23} \mu_{-31}, \quad
(\mu_+^{-1} \mu_-)_{22} = I_{m_2} - \mu_{+33} \mu_{-32}, \\
&&(\mu_+^{-1} \mu_-)_{23} = - \mu_{+23}, \quad 
(\mu_+^{-1} \mu_-)_{31} = \mu_{-31}, \\
&&(\mu_+^{-1} \mu_-)_{32} = \mu_{-32}, \quad
(\mu_+^{-1} \mu_-)_{33} = I_{m_3}.
\end{eqnarray*}
Now, using relations (\ref{a.10})--(\ref{a.13}) one can write down
the expressions for the mappings $\eta_{11}$, $\eta_{22}$ and $\eta_{33}$
entering the parametrisation of the mapping $\eta$,
\[
\eta = \left( \begin{array}{ccc}
\eta_{11} & 0 & 0\\
0 & \eta_{22} & 0\\
0 & 0 & \eta_{33}
\end{array} \right).
\]
The corresponding expressions are rather cumbersome, so we give here only 
one for $\eta_{11}$,
\[
\eta_{11} = I_{m_1} - \mu_{+12} \mu_{-21} - (\mu_{+13} - \mu_{+12}
\mu_{+23}) \mu_{-31}.
\]
Finally, relation (\ref{2.8}) allows us to write the general solution
of Toda equations (\ref{3.16})--(\ref{3.18}) in an explicit form.
In particular, the expression for the mapping $\beta_1$ is
\begin{equation}
\beta_1 = \beta_{+1}^{-1} (I_{m_1} - \mu_{+12} \mu_{-21} - (\mu_{+13}
- \mu_{+12} \mu_{+23}) \mu_{-31}) \beta_{-1}. \label{3.27}
\end{equation}

Consider now the case $m_1 = m_2 = m_3 = m$. Recall that in this
case the ${\Bbb Z}$-gradation under consideration can be associated
with the ${\frak sl}(2, {\Bbb C})$-subalgebra generated by the
elements $h$ and $x_\pm$ defined by (\ref{3.25}) and (\ref{3.26}).
Choosing again $c_\pm = x_\pm/\sqrt{2}$, we come to the Toda
equations (\ref{3.22})--(\ref{3.24}). Denote $\beta = \beta_1$, and
write $\beta$ in the following form
\[
\beta(z^-, z^+) = \zeta_{+1}(z^+) \zeta_{-1}(z^-) + \zeta_{+2}(z^+)
\zeta_{-2} (z^-) + \zeta_{+3}(z^+) \zeta_{-3}(z^-)
\]
which follows from (\ref{3.27}). Here
\bgroup \def\arraystretch{1.4} \[
\begin{array}{lcl}
\zeta_{+1} = \beta_{+1}^{-1}, &\qquad& \zeta_{-1} = \beta_{-1}, \\
\zeta_{+2} = -\beta_{+1}^{-1} \mu_{+12}, &\qquad& \zeta_{-2} =
\mu_{-21} \beta_{-1}, \\
\zeta_{+3} = - \beta_{+1}^{-1} (\mu_{+13} - \mu_{+12} \mu_{+23}),
&\qquad& \zeta_{-3} = \mu_{-31} \beta_{-1}.
\end{array}
\] \egroup
Using  equations (\ref{3.22}) and (\ref{3.23}), one obtains
\[
\beta_1 = \Delta_{1}, \qquad \beta_2 = - \Delta_{2}, \qquad
\beta_3 = \Delta_{3},
\]
where the quasideterminants $\Delta_{i}$ are defined by (\ref{3.12}).

It is interesting to go further and  consider the case when $r+1 =
p\, m$, where $p$ and $m$ are positive integers. Suppose that
$s_m = s_{2m} = \cdots = s_{(p-1)m} = 1$,
and all remaining labels of the Dynkin diagram are equal to zero. The
corresponding grading operator can be written as a $p \by p$ block
matrix $q = (q_{ij})$ with
\[
q_{ij} = \frac{1}{2} (p + 1 - 2i) \delta_{ij} I_m.
\]
The corresponding ${\Bbb Z}$-gradation of the Lie algebra ${\frak
gl}(r+1, {\Bbb C})$ can be associated with the ${\frak sl}(2, {\Bbb
C})$-subalgebra generated by the elements $h = (h_{ij})$, $x_- =
(x_{-ij})$ and $x_+ = (x_{+ij})$, where
\begin{eqnarray*}
&&h_{ij} = (p +1 - 2i) \delta_{ij} I_m, \\ 
&&x_{-ij} = \sqrt{i(p-i)} \delta_{i, j+1} I_m, \quad x_{+ij} =
\sqrt{i(p-i)} \delta_{i+1, j} I_m.
\end{eqnarray*}
Defining the elements $c_-$ and $c_+$ by the relations
\[
c_{-ij} = \delta_{i, j+1} I_m, \qquad c_{+ij} = \delta_{i+1,j} I_m,
\]
one comes to the following Toda equations
\begin{eqnarray}
&&\partial_+(\beta^{-1}_1 \partial_- \beta_1) = - \beta^{-1}_1
\beta_2, \label{3.30} \\
&&\partial_+(\beta^{-1}_i \partial_- \beta_i) = - \beta^{-1}_i
\beta_{i+1} + \beta_{i-1}^{-1} \beta_i, \quad 1 < i < p, \label{3.31}
\\
&&\partial_+(\beta^{-1}_p \partial_- \beta_p) = \beta^{-1}_{p-1}
\beta_p. \label{3.32}
\end{eqnarray}
Denote $\beta = \beta_1$, then one can show that from equations
(\ref{3.30})--(\ref{3.32}) it follows that
\begin{equation}
\beta_i = (-1)^{i-1} \Delta_{i}, \label{3.28}
\end{equation}
where the quasideterminants $\Delta_{i}$ are defined by
(\ref{3.12}). On the other hand, analysing the structure of the
Gauss decomposition of block matrices, we conclude that $\beta$ can
be represented as
\begin{equation}
\beta(z^-, z^+) = \sum_{i=1}^p \zeta_{+i}(z^+) \; \zeta_{-i}(z^-),
\label{3.33}
\end{equation}
where the mappings $\zeta_{\pm i}$ take values in the space of $m \by
m$ matrices. Using arbitrary mappings $\zeta_{\pm i}$, we obtain the
general solution to system (\ref{3.30})--(\ref{3.32}). A proof of
relation (\ref{3.28}) is given in appendix B.

Slightly modifying the consideration given in paper \cite{GRe92},
one can state the following. Consider the one dimensional
infinite system of ordinary differential equations
\begin{eqnarray*}
&&\frac{d}{dt} \left( \beta^{-1}_1 \frac{d
\beta_1}{dt} \right) = - \beta^{-1}_1 \beta_2, \\
&&\frac{d}{dt} \left( \beta^{-1}_i \frac{d
\beta_i}{dt} \right) = - \beta^{-1}_i
\beta_{i+1} + \beta_{i-1}^{-1} \beta_i, \quad i > 1,
\end{eqnarray*}
where $\beta_i$, $i = 1, 2, \ldots$, are ${\rm GL}(m, {\Bbb C})$ valued
functions of the real variable $t$. Denoting $\beta = \beta_1$,
one can write the general solution to this system as
\begin{equation}
\beta_i = (-1)^{i-1} \Gamma_{i}, \label{3.29}
\end{equation}
where $\Gamma_{i}$, $i=1,2,\ldots$, are the quasideterminants defined
by
\[
\Gamma_{i} = \left| \begin{array}{cccc}
\beta & \displaystyle \frac{d \beta}{dt} & \cdots & \displaystyle
\frac{d^{i-1} \beta}{dt^{i-1}} \\[.6em]
\displaystyle \frac{d \beta}{dt} & \displaystyle \frac{d^2
\beta}{dt^2} & \cdots & 
\displaystyle \frac{d^i \beta}{dt^i} \\[.6em]
\vdots \hfil & \vdots \hfil & \ddots & \vdots \hfil \\[.6em]
\displaystyle \frac{d^{i-1} \beta}{dt^{i-1}} & \displaystyle
\frac{d^i \beta}{dt^i} & \cdots &
\displaystyle \frac{d^{2i-2} \beta}{dt^{2i-2}}
\end{array} \right|_{ii}. 
\]
Hence, relation (\ref{3.28}) can be considered as a two dimensional
generalisation of (\ref{3.29}). Actually, relation (\ref{3.28}) for
arbitrary $\beta$ gives the general solution to the infinite system
of equations having form (\ref{3.30}) and (\ref{3.31}) without the
condition $i < p$. Here if $\beta$ has form (\ref{3.33}),
then one gets $\Delta_{p+1} = 0$ and comes to the general
solution of the finite system (\ref{3.30})--(\ref{3.32}). Note that the
expression for $\beta_1$ in form $\beta_1 = \beta$ with $\beta$ given
by (\ref{3.33}), was also obtained in paper \cite{LYu95} by some other
method. 

\section{Complex orthogonal group}

The complex orthogonal group ${\rm O}(n, {\Bbb C})$ is the Lie
subgroup of the Lie group ${\rm GL}(n, {\Bbb C})$ formed by matrices
$a \in {\rm GL}(n, {\Bbb C})$ satisfying the condition 
\begin{equation}
\tilde I_n a^t \tilde I_n = a^{-1}, \label{4.11}
\end{equation} 
where $\tilde I_n$ is the antidiagonal unit $n \by n$ matrix, and
$a^t$ is the transpose of $a$.  The corresponding Lie algebra ${\frak
o}(n, {\Bbb C})$ is the subalgebra of ${\frak gl}(n, {\Bbb C})$ which
consists of the matrices $x$ satisfying the condition
\begin{equation}
\tilde I_n x^t \tilde I_n = -x. \label{4.12}
\end{equation}
For an $m_1 \by m_2$ matrix $a$ we will denote by $a^T$ the matrix
defined by the relation
\[
a^T = I_{m_2} a^t I_{m_1}.
\]
Using this notation, we can rewrite conditions (\ref{4.11}) and
(\ref{4.12}) as $a^T = a^{-1}$ and $x^T = -x$.  The Lie algebra
${\frak o}(n, {\Bbb C})$ is simple.  For $n = 2r+1$ it is of type
$B_r$, while for $n = 2r$ it is of type $D_r$. Discuss these two
cases separately.

Consider the ${\Bbb Z}$-gradation of ${\frak o}(2r+1, {\Bbb C})$
arising when we choose $s_d = 1$ for some fixed $d$ such that $1 \le
d \le r$, and put all other labels of the Dynkin diagram be equal to
zero.  Using relation (\ref{2.5}), one gets
\begin{eqnarray*}
&&q = \sum_{i=1}^{r-1} i h_i + \frac{1}{2} r h_r, \quad d = r, \\
&&q = \sum_{i=1}^{r-1} i h_i + \frac{1}{2} (r-1) h_r, \quad d = r-1, \\
&&q = \sum_{i=1}^d i h_i + d \sum_{i=d+1}^{r-1} h_i + \frac{1}{2} d
h_r, \quad 1 \le d < r-1.  
\end{eqnarray*}
It is convenient to choose the following Cartan generators of ${\frak
o}(2r+1, {\Bbb C})$:
\begin{eqnarray*}
&&h_i = e_{i, i} - e_{i+1, i+1} + e_{2r+1-i, 2r+1-i} - e_{2r+2-i,
2r+2-i}, \qquad 1 \le i < r, \\
&&h_r = 2(e_{r,r} - e_{r+2, r+2}),
\end{eqnarray*}
where the matrices $e_{i,j}$ are defined by (\ref{3.34}). Using these
expressions one obtains
\[
q = \sum_{i=1}^d e_{i, i} - \sum_{i=1}^d e_{2r+2-i, 2r+2-i}. 
\]
Denoting $m_1 = d$ and $m_2 = 2(r-d) + 1$, we write $q$ in the block
matrix form,
\begin{equation}
q = \left( \begin{array}{ccc}
I_{m_1} & 0 & 0 \\
0 & 0 & 0 \\
0 & 0 & -I_{m_1}
\end{array} \right), \label{4.3}
\end{equation}
where zero on the diagonal stands for the $m_2 \times m_2$ block of
zeros.

It is easy to verify that in the case of the Lie algebra ${\frak
o}(2r, {\Bbb C})$ 
\begin{eqnarray*}
&&q = \frac{1}{2} \sum_{i=1}^{r-2} i h_i + \frac{1}{4} (r-2) h_{r-1} +
\frac{1}{4} r h_r, \quad d = r, \\
&&q = \frac{1}{2} \sum_{i=1}^{r-2} i h_i + \frac{1}{4} r h_{r-1} +
\frac{1}{4} (r-2) h_r, \quad d = r-1, \\
&&q = \sum_{i=1}^d i h_i + d \sum_{i = d+1}^{r-2} h_i + \frac{1}{2} d
(h_{r-1} + h_r), \quad 1 \le d < r-1. 
\end{eqnarray*}
Choose as the Cartan generators of ${\frak o}(2r, {\Bbb C})$ the
elements 
\begin{eqnarray*}
&&h_i = e_{i,i} - e_{i+1, i+1} + e_{2r-i, 2r-i} - e_{2r+1-i, 2r+1-i},
\quad 1 \le i < r, \\
&&h_r = e_{r-1, r-1} + e_{r,r} - e_{r+1, r+1} - e_{r+2, r+2}.
\end{eqnarray*}
Then one easily obtains 
\begin{eqnarray*}
&&q = \frac{1}{2} \sum_{i=1}^r e_{i, i} - \frac{1}{2} \sum_{i=1}^r
e_{2r+1-i, 2r+1-i}, \quad d = r, \\
&&q = \frac{1}{2} \sum_{i=1}^{r-1} e_{i, i} - \frac{1}{2} e_{r, r} +
\frac{1}{2} e_{r+1, r+1} - \frac{1}{2} \sum_{i=1}^{r-1} e_{2r+1-i,
2r+1-i}, \quad d = r-1, \\
&&q = \sum_{i=1}^d e_{i, i} - \sum_{i=1}^d e_{2r+1-i, 2r+1-i}, \quad
1 \le i < r-1. 
\end{eqnarray*}
Note that the grading operators corresponding to the cases $d = r$
and $d = r-1$ are connected by the automorphism $\sigma$ of ${\frak
o}(2r, {\Bbb C})$ defined by the relation $\sigma(x) = axa^{-1}$,
where $a$ is the matrix corresponding to the permutation of the
indices $r$ and $r+1$.  There is the corresponding automorphism of
the Lie group ${\rm O}(2r, {\Bbb C})$, which is defined by the same
formula. Thus, the cases $d=r$ and $d=r-1$ leads actually to the same
Toda equations, and we can exclude one of them, for example $d=r-1$
from the consideration.

For the case $d=r$ the grading operator has the following block form
\begin{equation}
q = \frac{1}{2} \left( \begin{array}{cc}
I_m & 0 \\
0 & -I_m
\end{array} \right), \label{4.4}
\end{equation}
where we denoted $m = r$. In the case $1 \le d < r-2$ denoting $m_1 =
d$ and $m_2 = 2(r-d)$ one sees that the grading operator $q$ has form
(\ref{4.3}). 

Resuming the consideration, we can say that for any representation of
the positive integer $n$ in the form $n = 2m_1 + m_2$ where $m_1$
and $m_2$ are positive integers such that $m_2 \ne 2$, there is the
${\Bbb Z}$-gradation of ${\frak o}(n, {\Bbb C})$ corresponding to a
maximally nonabelian Toda system. This gradation is generated by the
grading operator (\ref{4.3}). In the case $n = 2m$, there is one more
${\Bbb Z}$-gradation defined by the grading operator (\ref{4.4}).

Probably, the absence of the case $m_2 = 2$ requires a special
explanation. Actually the operator $q$ given by (\ref{4.3}) with $m_2
= 2$ defines some ${\Bbb Z}$-gradation of the corresponding complex
orthogonal algebra, but this gradation corresponds to the case when
two labels of the Dynkin diagram, $s_{r-1}$ and $s_r$, are equal to 1.
Therefore, the corresponding Toda system is not maximally abelian.
Nevertheless, it is convenient to consider the case $m_2 = 2$ together
with those which do correspond to maximally nonabelian systems.

The ${\Bbb Z}$-gradation defined by the grading operator (\ref{4.4})
can be associated with an ${\rm SL}(2, {\Bbb C})$-subalgebra of the
Lie algebra ${\frak o}(2m, {\Bbb C})$ if and only if the integer $m$ is
even. In this case the corresponding element $h$ coincides with $2q$,
and the elements $x_\pm$ have form (\ref{3.36}), where the matrices
$X_\pm$ satisfy the conditions
\[
X_+ X_- = I_m, \qquad X_\pm^T = - X_\pm.
\]
With a ${\Bbb Z}$-gradation generated by the grading operator
of form (\ref{4.3}) one can find the corresponding ${\rm SL}(2, {\Bbb
C})$-subalgebra of ${\frak o}(n, {\Bbb C})$ if and only if $m_1 \le
m_2$. Here $h = 2q$ and the elements $x_\pm$ are
\[
x_- = \left( \begin{array}{ccc}
0 & 0 & 0 \\
X_- & 0 & 0 \\
0 & -X_-^T & 0
\end{array} \right), \qquad 
x_+ = \left( \begin{array}{ccc}
0 & X_+ & 0 \\
0 & 0 & -X_+^T \\
0 & 0 & 0
\end{array} \right),
\]
where the matrices $X_\pm$ satisfy the conditions
\[
X_+ X_- = 2 I_{m_1}, \qquad X_+^T X_-^T - X_- X_+ = 0.
\]

Proceed now to the consideration of the Toda equations associated to
the gradations described above. Begin with the gradation
generated by the grading operator (\ref{4.4}). The general form of
the elements $c_\pm \in {\frak g}_{\pm 1}$ is given by (\ref{3.35}),
where the matrices $C_\pm$ should satisfy the relations
\begin{equation}
C_\pm^T = - C_\pm. \label{4.5}
\end{equation}
The subgroup $\tilde H$ in the case under consideration is
formed by the $2 \by 2$ block matrices of the form
\[
h = \left( \begin{array}{cc}
h_{11} & 0 \\
0 & (h_{11}^{-1})^T
\end{array} \right),
\]
where $h_{11} \in {\rm GL}(m, {\Bbb C})$. Hence, the mapping
$\gamma$ has the following block form
\begin{equation}
\gamma = \left( \begin{array}{cc}
\beta & 0 \\
0 & (\beta^{-1})^T
\end{array} \right), \label{4.15}
\end{equation}
where the mapping $\beta$ take values in the Lie group ${\rm GL}(m,
{\Bbb C})$. The corresponding Toda equations are
\begin{equation}
\partial_+(\beta^{-1} \partial_- \beta) = -\beta^{-1} C_+
(\beta^{-1})^T C_-. \label{4.6}
\end{equation}
It is clear that these equations can be considered as the result of
the reduction of equations (\ref{3.2}), (\ref{3.3}) to the case
$\beta_2 = (\beta_1^{-1})^T$, which is possible if relations
(\ref{4.5}) are valid. Therefore, the general solution to equation
(\ref{4.6}) can be obtained from the general solution to equations
(\ref{3.2}), (\ref{3.3}) by the method described in \cite{RSa96a}.

Certainly, we can get the general solution directly, using the
general scheme of section \ref{int}. The mappings $\gamma_\pm$ in the
case under consideration have the form
\[
\gamma_\pm = \left( \begin{array}{ccc}
\beta_\pm & 0 \\
0 & (\beta_\pm^{-1})^T
\end{array} \right),
\]
where the mappings $\beta_\pm$ take values in the Lie group ${\rm
GL}(m, {\Bbb C})$.  Note that the subgroups $\tilde N_-$ and
$\tilde N_+$ in the case under consideration are formed
respectively by the block matrices
\[
n_- = \left( \begin{array}{cc}
I_m & 0 \\
n_{-21} & I_m
\end{array} \right), \qquad
n_+ = \left( \begin{array}{cc}
I_m & n_{+12} \\
0 & I_m
\end{array} \right),
\]
where the matrices $n_{-21}$ and $n_{+12}$ obey the equalities
\[
n_{-21}^T = - n_{-21}, \qquad n_{+12}^T = - n_{+12}.
\]
Representing the mappings $\mu_\pm$ as
\[
\mu_- = \left( \begin{array}{cc}
I_m & 0 \\
\mu_{-21} & I_m
\end{array} \right), \qquad
\mu_+ = \left( \begin{array}{cc}
I_m & \mu_{+12} \\
0 & I_m
\end{array} \right), 
\]
one obtains
\begin{eqnarray}
&&\mu_{-21} (z^-) = m_{-21} + \int_0^{z^-} dy^- (\beta_-^{-1})^T (y^-) C_-
\beta_{-}^{-1} (y^-), \label{4.7} \\ 
&&\mu_{+12} (z^+) = m_{+12} + \int_0^{z^+} dy^+ \beta_+ (y^+) C_+
\beta_+^T (y^+), \label{4.8}
\end{eqnarray}
where the constant matrices $m_{-21}$ and $m_{+12}$ satisfy the relations
\[
m_{-21}^T = - m_{-21}, \qquad m_{+12}^T = - m_{+12}.
\]
The final expression for the general solution to equations
(\ref{4.6}) is
\begin{equation}
\beta = \beta_+^{-1} (I_m - \mu_{+12} \mu_{-21}) \beta_-, \label{4.16}
\end{equation}
where the mappings $\mu_{-21}$ and $\mu_{+12}$ are given by
(\ref{4.7}) and (\ref{4.8}).

Consider now the Toda equations arising when we choose the ${\Bbb
Z}$-gradation of ${\frak o}(n, {\Bbb C})$ generated by the grading
operator $q$ defined by (\ref{4.3}). In this case the general form of
the elements $c_\pm$ is
\[
c_- = \left( \begin{array}{ccc}
0 & 0 & 0 \\
C_- & 0 & 0 \\
0 & - C_-^T & 0
\end{array} \right), \qquad 
c_+ = \left( \begin{array}{ccc}
0 & C_+ & 0 \\
0 & 0 & - C_+^T \\
0 & 0 & 0 
\end{array} \right).
\]
The mapping $\gamma$ has the following block form
\begin{equation}
\gamma = \left( \begin{array}{ccc}
\beta_1 & 0 & 0\\
0 & \beta_2 & 0\\ 
0 & 0 & (\beta_1^{-1})^T
\end{array} \right), \label{4.17}
\end{equation}
where the mappings $\beta_1$ and $\beta_2$ take values in the Lie
groups ${\rm GL}(m_1, {\Bbb C})$ and ${\rm O}(m_2, {\Bbb C})$
respectively. The corresponding Toda equations are
\begin{eqnarray*}
&&\partial_+(\beta^{-1}_1 \partial_- \beta_1) = - \beta^{-1}_1 C_+
\beta_2 C_-, \label{4.9} \\
&&\partial_+(\beta^{-1}_2 \partial_- \beta_2) = - 
(C_- \beta_1^{-1} C_+ \beta_2)^T
+ C_- \beta_1^{-1} C_+ \beta_2. \label{4.10}
\end{eqnarray*}
These equations can be considered as the reduction of equations
(\ref{3.16})--(\ref{3.18}) to the case $\beta_3 = \beta_1$ and
$\beta_2^T = \beta_2^{-1}$. Such a reduction is possible if the
matrices $C_{\pm 1}$ and $C_{\pm 2}$ in (\ref{3.16})--(\ref{3.18})
satisfy the conditions
\[
C_{-1} = - C_{-2}^T = C_-, \qquad C_{+1} = -C_{+2}^T = C_+.
\]
The general solution to equations (\ref{4.9}), (\ref{4.10}) can be
obtained either by the reduction of the general solution to equations
(\ref{3.16})--(\ref{3.18}), or directly. The corresponding
expressions are rather cumbersome and we do not give them here. Note
only that the corresponding subgroups $\tilde N_-$ and
$\tilde N_+$ are composed respectively from the block matrices of
the form
\begin{equation}
n_- = \left( \begin{array}{ccc}
I_{m_1} & 0 & 0 \\
n_{-21} & I_{m_2} & 0 \\
n_{-31} & n_{-32} & I_{m_1}
\end{array} \right), \qquad
n_+ = \left( \begin{array}{ccc}
I_{m_1} & n_{+12} & n_{+13} \\
0 & I_{m_2} & n_{+23} \\
0 & 0 & I_{m_1}
\end{array} \right). \label{4.18}
\end{equation}
where
\begin{eqnarray*}
&&n_{+23}^T = - n_{+12}, \qquad n_{+13}^T = -n_{+13} + n_{+12}
n_{+23}, \\
&&n_{-31}^T = - n_{-21}, \qquad n_{-31}^T = -n_{-31} + n_{-32}
n_{-21}. 
\end{eqnarray*}

The maximally nonabelian Toda system based on the Lie group ${\rm
O}(5, {\Bbb C})$ and with the choice $d=1$ in relation to the
physics of black holes was investigated in paper \cite{GSa92}, see
also \cite{Bil94}. In these papers there was used a local
parametrisation of the Lie group $\tilde H$, which is isomorphic
here to ${\rm GL}(1, {\Bbb C}) \times {\rm O}(3, {\Bbb C})$. With
the approach developed in the present paper, one can write the general
solution in terms of corresponding matrices without using any local
coordinates. Actually, for the system considered in \cite{GSa92} and
\cite{Bil94} it is easier to use the fact that the Lie group
${\rm O}(5, {\Bbb C})$ is locally isomorphic to the Lie group ${\rm
Sp}(4, {\Bbb C})$ and to consider the corresponding maximally
nonabelian Toda system based on ${\rm Sp}(4, {\Bbb C})$. It will be
done in the next section. 

\section{Complex symplectic group}

We define the complex symplectic group ${\rm Sp}(2r, {\Bbb C})$ as the
Lie subgroup of the Lie group ${\rm GL}(2r, {\Bbb C})$ which consists
of the matrices $a \in {\rm GL}(2r, {\Bbb C})$ satisfying the
condition
\[
\tilde J_r a^t \tilde J_r  = - a^{-1},
\]
where $\tilde J_r$ is the matrix given by
\[
\tilde J_r = \left( \begin{array}{cc}
0 & \tilde I_r \\
-\tilde I_r & 0
\end{array} \right).
\]
The corresponding Lie algebra ${\frak sp}(r, {\Bbb C})$ is defined as
the subalgebra of the Lie algebra ${\frak sl}(2r, {\Bbb C})$ formed
by the matrices $x$ which satisfy the condition
\[
\tilde J_r x^t \tilde J_r = x.       
\]
The Lie algebra ${\frak sp}(r, {\Bbb C})$ is simple, and it is of
type $C_r$. Therefore, the Cartan matrix of ${\frak sp}(r, {\Bbb C})$
is the transpose of the Cartan matrix of ${\frak o}(n, {\Bbb C})$,
and the same is for inverse of the Cartan matrix of ${\frak sp}(r,
{\Bbb C})$. For any fixed integer $d$ such that $1 \le d \le r$,
consider the ${\Bbb Z}$-gradation of ${\frak sp}(r, {\Bbb C})$
arising when we choose all the labels of the corresponding Dynkin
diagram equal to zero, except the label $s_d$, which we choose to be
equal to $1$. Using relation (\ref{2.5}), we obtain the following
expressions for the grading operator,
\[
q = \frac{1}{2} \sum_{i=1}^r i h_i, \quad d = r, \qquad
q = \sum_{i=1}^d i h_i + d \sum_{i=d+1}^r h_i, \quad 1 \le d < r.
\]
Using the following choice of the Cartan generators,
\begin{eqnarray*}
&&h_i = e_{i,i} - e_{i+1, i+1} + e_{2r-i, 2r-i} - e_{2r+1-i, 2r+1-i},
\qquad 1 \le i < d, \\
&&h_r = e_{r, r} - e_{r+1, r+1},
\end{eqnarray*}
one sees that the grading operator for the case $d = r$ has form
(\ref{4.4}) with $m = r$, and for the case $1 \le d < r$ it has form
(\ref{4.3}) with $m_1 = d$ and $m_2 = 2(r-d)$. The corresponding
${\rm SL}(2, {\Bbb C})$-subalgebra always exists for the case $d=r$,
and for the case $1 \le d < r$ it exists if and only if $m_1 \le
m_2$.

In the case $d = r$ the general form of the elements $c_\pm$ is given
by (\ref{3.35}), where the matrices $C_\pm$ satisfy the conditions
\begin{equation}
C_\pm^T = C_\pm. \label{5.1}
\end{equation}
The mapping $\gamma$ has here form (\ref{4.15}) and the Toda
equations coincide with (\ref{4.6}) where the matrices $C_\pm$
satisfy (\ref{5.1}). The obtained equations can be considered as the
reduction of  equations (\ref{3.2}), (\ref{3.3}) to the case
$\beta_1 = (\beta_2^{-1})^T = \beta$ which is possible when
(\ref{5.1}) is valid.  The general solution of the Toda equations is
described by relation (\ref{4.16}) where the mappings $\mu_{-21}$
and $\mu_{+12}$ are given by (\ref{4.7}) and (\ref{4.8}) with the
constant matrices $m_{-21}$ and $m_{+12}$ satisfying the relations
\[
m_{-21}^T = m_{-21}, \qquad m_{+12}^T = m_{+12}.
\]

The simplest choice of the matrices $C_\pm$ is $C_\pm = I_r$. Here
the Toda equations take the form
\[
\partial_+ (\beta^{-1} \partial_- \beta) = - (\beta^T \beta)^{-1}. 
\]
The subgroups $\tilde H_\pm$ determining the symmetry
transformations (\ref{2.13}) are isomorphic to the Lie group ${\rm
O}(r, {\Bbb C})$ and are composed of the matrices of form
(\ref{3.37}), where the matrix $h_{11}$ satisfy the condition
$h_{11}^T = h_{11}^{-1}$. With $C_\pm = \tilde I_r$ we
come to the equations
\[
\partial_+ (\beta^{-1} \partial_- \beta) = - (\beta^t \beta)^{-1}.
\]

Return to the discussion given in the end of the previous
section. It is clear that choosing consistent local parametrisation
for the Lie groups ${\rm Sp}(4, {\Bbb C})$ and ${\rm O}(5, {\Bbb
C})$, we can obtain from the general solution of the Toda equations for
${\rm Sp}(4, {\Bbb C})$ the general solution for the corresponding
Toda equations based on ${\rm O}(5, {\Bbb C})$. It can be verified
that this solution coincides with the solution obtained in
\cite{Bil94}.

Proceed now to the case $1 \le d < r$. In this case the general form
of the elements $c_\pm$ is
\[
c_- = \left( \begin{array}{ccc}
0 & 0 & 0 \\
C_- & 0 & 0 \\
0 & - \tilde I_d C_-^t \tilde J_{r-d} & 0
\end{array} \right), \qquad
c_+ = \left( \begin{array}{ccc}
0 & C_+ & 0 \\
0 & 0 & \tilde J_{r-d} C_+^t \tilde I_d \\
0 & 0 & 0
\end{array} \right);
\]
the mapping $\gamma$ has form (\ref{4.17}) where the mappings
$\beta_1$ and $\beta_2$ take values in the Lie groups ${\rm GL}(d,
{\Bbb C})$ and ${\rm Sp}(2(r-d), {\Bbb C})$ respectively; and the Toda
equations are
\begin{eqnarray}
&&\partial_+ (\beta_1^{-1} \partial_- \beta_1) = - \beta_1^{-1} C_+
\beta_2 C_-, \label{5.2} \\
&&\partial_+ (\beta_2^{-1} \partial_- \beta_2) = \beta_2^{-1}
\tilde J_{r-d} C_+^t (\beta_1^{-1})^t C_-^t \tilde J_{r-d} +
C_- \beta_1^{-1} C_+ \beta_2. \label{5.3}
\end{eqnarray}
These equations are the reduction of equations
(\ref{3.16})--(\ref{3.18}) to the case $\beta_3 = \beta_1$ and
$\tilde J_{r-d} \beta_2^t \tilde J_{r-d} = - \beta_2^{-1}$,
which is possible if the matrices $C_{\pm 1}$ and $C_{\pm 2}$ in
(\ref{3.16})--(\ref{3.18}) satisfy the conditions
\[
C_{-1} = - \tilde J_{r-d} C_{-2}^t \tilde I_d = C_-, \qquad
C_{+1} = \tilde I_d C_{+2}^t \tilde J_{r-d} = C_+.
\]
It is quite clear that using the integration procedure described in
section \ref{int}, one can easily write the general solution for 
equations (\ref{5.2}), (\ref{5.3}).

\section{Concluding remarks}

The main goal of our study was to obtain explicit expressions for the
general solutions for some class of nonabelian Toda systems, namely
for the maximally nonabelian ones, which has a very simple structure.
Our consideration concerns finite dimensional complex semisimple Lie
groups, but can be extended for the exceptional groups and infinite
dimensional loop groups. In particular, starting with the loop group
associated with the complex general group it is possible to come to
the periodic two dimensional nonabelian Toda chain which was obtained
by H.~W.~ Capel and J.~H.~H.~Perk in the context of the inhomogeneous
$XY$--model \cite{PCa78}, and  also by A.~V.~Mikhailov in the
framework of the reduction scheme \cite{Mik81}. Finite-gap solutions
to this equations were found by I.~M.~Krichever \cite{Kri81}.

Actually one can consider the infinite system of equations of form
(\ref{3.31}) with $i = 0, \pm 1, \pm 2, \ldots$. This system was
introduced by A.~M.~Polyakov at the end of seventies. The finite two
dimensional nonabelian Toda chain and the periodic one can be
considered as special cases of such system corresponding to the
boundary conditions $\beta_0^{-1} = \beta_{p+1} = 0$ and $\beta_0 =
\beta_{p+1}$ respectively.

We believe that nonabelian Toda systems are quite relevant for a
number of applications in theoretical and mathematical physics,
especially in particle and statistical physics, and in a near future
their role for the description of nonlinear phenomena in these areas
will be not less than that for the abelian systems.

Finishing up the paper, we would like to mention about one more
reason why nonabelian Toda systems are not yet popular and known
enough in mathematics, in particular in algebraic and differential
geometry. The problem is that such important notions as the Gauss-Manin
flat connection, the Griffiths transversality of variations of the
Hodge structures, superhorizontal distributions, local and global
Pl\"ucker relations, etc., which are perfectly fitted in the scheme
with abelian Toda system, need to be re-understood or extended for
nonabelian case. The same is for the systems generated  by flat
connections with values in higher grading subspaces of complex ${\Bbb
Z}$-graded Lie algebra.  We hope that the progress in this direction
will be very fruitful for studies in theoretical physics as well as
for mathematics itself.

\section*{Acknowlegements}

The authors are grateful to J.-L.~Gervais, I.~M.~Krichever,
Yu.~I.~Manin and J.~H.~H.~Perk for the fruitful discussions.  One of
the authors (M.~V.~S.) wishes to acknowledge the warm hospitality and
creative scientific atmosphere of the Laboratoire de Physique
Th\'eorique de l'\'Ecole Normale Sup\'erieure de Paris.

\section*{Note added in proof}

After our paper was submitted to the journal there appeared an
electronic preprint of P.~Etingof, I.~Gelfand and V.~Retakh
\cite{EGR97} concerning in particular nonabelian Toda systems. In
this interesting paper some of our results are reproduced by some other
method.

\appendix

\def\theequation{A.\arabic{equation}}

\section{Gauss decomposition of block matrices}

Let $a = (a_{ij})$ be a nondegenerate block matrix formed by $m_i \by m_j$
matrices. By the Gauss decomposition of such a matrix we mean its
representation as the product of a lower triangular block matrix with
the unit matrices on the diagonal, an upper triangular block matrix
with the unit matrices on the diagonal, and a block diagonal matrix,
taken in some order. In this appendix we construct the Gauss
decomposition of $2 \by 2$ and $3 \by 3$ matrices.

Consider first a nondegenerate $2 \by 2$ block matrix
\begin{equation}
a = \left(\begin{array}{cc}
a_{11} & a_{12} \\
a_{21} & a_{22}
\end{array} \right). \label{a.2}
\end{equation}
Suppose that the matrix $a$ can be represented as
\begin{equation}
a = n_- h n_+^{-1}, \label{a.1}
\end{equation}
where $n_-$, $h$ and $n_+$ are block matrices of the form
\[
n_- = \left( \begin{array}{cc}
I_{m_1} & 0 \\
n_{-12} & I_{m_2}
\end{array} \right), \quad
h = \left( \begin{array}{cc}
h_{11} & 0 \\
0 & h_{22}
\end{array} \right), \quad
n_+ = \left( \begin{array}{cc}
I_{m_1} & n_{+12} \\
0 & I_{m_2}
\end{array} \right).
\]
Note that since the matrix $a$ is nondegenerate, the matrices $h_{11}$
and $h_{22}$ must be also nondegenerate. It can be easily shown that
\[
n_+^{-1} = \left( \begin{array}{cc}
I_{m_1} & - n_{+12} \\
0 & I_{m_2}
\end{array} \right).
\]
Using this relation, one can write equality (\ref{a.1}) in the
form 
\[
\left( \begin{array}{cc}
a_{11} & a_{12} \\
a_{21} & a_{22}
\end{array} \right) =
\left( \begin{array}{cc}
h_{11} & - h_{11} n_{+12} \\
n_{21} h_{11} & h_{22} - n_{-21} h_{11} n_{+12}
\end{array} \right).
\]
{}From this equality it follows that the Gauss decomposition of the
matrix $a$ of form (\ref{a.2}) exists if and only if the matrix
$a_{11}$ is nondegenerate. In this case one has
\begin{eqnarray}
&&n_{-21} = a_{21} (a_{11})^{-1}, \label{a.3} \\
&&h_{11} = a_{11}, \label{a.4} \\
&&h_{22} = a_{22} - a_{21} (a_{11})^{-1} a_{12}, \label{a.5} \\
&&n_{+12} = - (a_{11})^{-1} a_{12}. \label{a.6}
\end{eqnarray}
It is worth to note that the obtained Gauss decomposition is unique.

Consider now a nondegenerate $3 \by 3$ matrix
\[
a = \left(\begin{array}{ccc}
a_{11} & a_{12} & a_{13} \\
a_{21} & a_{22} & a_{23} \\
a_{31} & a_{32} & a_{33}
\end{array} \right).
\]
Suppose that $a$ can be represented in form (\ref{a.1}), where
\begin{eqnarray*}
&h = \left(\begin{array}{ccc}
h_{11} & 0 & 0 \\
0 & h_{22} & 0 \\
0 & 0 & h_{33}
\end{array} \right), \\
&n_- = \left(\begin{array}{ccc}
I_{m_1} & 0 & 0 \\
n_{-21} & I_{m_2} & 0 \\
n_{-31} & n_{-32} & I_{m_3}
\end{array} \right), \qquad
n_+ = \left(\begin{array}{ccc}
I_{m_1} & n_{+12} & n_{+13} \\
0 & I_{m_2} & n_{+23} \\
0 & 0 & I_{m_3}
\end{array} \right).
\end{eqnarray*}
Now we have
\[
n_+^{-1} = \left( \begin{array}{ccc}
I_{m_1} & - n_{+12} & - n_{+13} + n_{+12} n_{+23} \\
0 & I_{m_2} & - n_{+23} \\
0 & 0 & I_{m_3}
\end{array} \right),
\]
and in the same way as it was done for the case of $2 \by 2$ block
matrices one finds out that the Gauss decomposition in question
exists if and only if the matrices $a_{11}$ and $a_{22} - a_{21}
(a_{11})^{-1} a_{21}$ are nondegenerate. The explicit expressions
determining the matrices $n_-$, $h$ and $n_+$ are
\begin{eqnarray}
&&n_{-21} = a_{21} (a_{11})^{-1}, \label{a.7} \\
&&n_{-31} = a_{31} (a_{11})^{-1}, \label{a.8} \\
&&n_{-32} = (a_{32} - a_{31} (a_{11})^{-1} a_{12}) (a_{22} - a_{21}
(a_{11})^{-1} a_{12})^{-1}, \label{a.9} \\
&&h_{11} = a_{11}, \label{a.10} \\
&&h_{22} = a_{22} - a_{21} (a_{11})^{-1} a_{12}, \label{a.11} \\
&&h_{33} = a_{33} - a_{31} (a_{11})^{-1} a_{13} 
+ (a_{32} - a_{31} (a_{11})^{-1} a_{12}) \nonumber \\
&& \hskip 7em \times (a_{22} -
a_{21} (a_{11})^{-1} a_{12})^{-1} (a_{23} - a_{21} (a_{11})^{-1}
a_{13}), \label{a.13} \\
&&n_{+12} = - (a_{11})^{-1} a_{12}, \label{a.14} \\
&&n_{+13} = - (a_{11})^{-1} (a_{13} - a_{12} (a_{22} - a_{21}
(a_{11})^{-1} a_{12})^{-1} \nonumber \\
&& \hskip 7em \times (a_{23} - a_{21} (a_{11})^{-1} a_{13})),
\label{a.15} \\ 
&&n_{+23} = - (a_{22} - a_{21} (a_{11})^{-1} a_{12})^{-1} (a_{23} -
a_{21} (a_{11})^{-1} a_{13}). \label{a.16}
\end{eqnarray}
The Gauss decomposition in this case is again unique.

It is interesting  to compare the relations obtained in this appendix
with those arising in the theory of quasideterminants of matrices over
an associative unital ring, in the form proposed by I. M. Gelfand and
V. S. Retakh \cite{GRe91,GRe92}, see also \cite{RSa96}. Let us give
here some relevant definitions.  

Let ${\cal I}$ and ${\cal J}$ be two ordered sets, each consisting of
$p$ elements. Consider an invertible matrix $a = (a_{ij})_{i \in
{\cal I}, j \in {\cal J}}$ with the matrix elements belonging to some
associative unital ring $R$. Define the family of $p^2$ elements
$|a|_{ij}$, $i \in {\cal I}$, $j \in {\cal J}$, of the ring $R$,
which are called the quasideterminants of $a$. In the case $p=1$
there is only one quasideterminant defined by $|a|_{ij} = a_{ij}$.
For $p > 1$ the quasideterminant $|a|_{ij}$ of the matrix $a$
can be defined by the relation
\begin{equation}
|a|_{ij}^{-1} = (a^{-1})_{ji}. \label{a.19}
\end{equation}
It is clear that the quasideterminant $|a|_{ij}$ exists only if the
matrix element $(a^{-1})_{ji}$ is an invertible element of the ring
$R$.  A more general definition of a quasideterminant applicable to
the case of noninvertible matrices can be found in
\cite{GRe92}, but for our purposes the above definition is most
convenient.  

Let ${\cal K}$ and ${\cal L}$ be subsets of ${\cal I}$ and ${\cal J}$
respectively. Denote by $a^{({\cal K}; {\cal L})}$ the matrix which is
obtained from the matrix $a$ by removing the matrix elements $a_{ij}$
with $i \in {\cal K}$ or $j \in {\cal L}$.  Consider the matrix
$a^{(i; j)}$ and suppose that it is invertible. It is not difficult
to show that
\begin{equation}
|a|_{ij} = a_{ij} - \sum_{k \ne i, l \ne j} a_{ik}
(a^{(i;j)-1})_{kl} a_{lj}. \label{a.18}
\end{equation}
This equality is used in appendix B to prove the validity of
relation (\ref{3.28}).

Further, denote by $a_{({\cal K}; {\cal L})}$ the submatrix of $a$
composed of the matrix elements $a_{ij}$ with $i \in {\cal K}$ and $j
\in {\cal L}$. Consider the case when ${\cal I} = {\cal J} =
\{1, \ldots, p\}$. It can be shown that an invertible matrix $a =
(a_{ij})_{i \in {\cal I}, j \in {\cal J}}$ has the Gauss
decomposition of form (\ref{a.1}) if and only if there exist the
quasideterminants $|a_{(1,\ldots,i;1,\ldots,i)}|_{ii}$, $i = 1,
\ldots, p$. Here the nontrivial matrix elements of the matrix $h$ are
determined by these quasideterminants. Actually, one has
\begin{equation}
h_{ii} = |a_{(1,\ldots,i;1,\ldots,i)}|_{ii}. \label{a.17}
\end{equation}

Returning to the case of block matrices, suppose that all blocks of
the matrices under consideration are square $m \by m$ matrices. In
this case we can treat such a block matrix as a matrix over the ring
of $m \by m$ matrices and use the formulae relevant for matrices over
the general associative unital ring.

\section{Proof of relation (3.33)}

\def\theequation{B.\arabic{equation}}

In this appendix we prove relation (\ref{3.28}) which gives the
general solution to equations (\ref{3.30})--(\ref{3.32}). Probably
our proof is not the shortest one but it is quite direct.

First of all note that from (\ref{3.31}) it follows that
\[
\beta_{i+1} = - \partial_+ \partial_- \beta_i + \partial_+ \beta_i
\beta_i^{-1} \partial_- \beta_i + \beta_i \beta_{i-1}^{-1} \beta_i.
\]
Therefore, to prove (\ref{3.28}) it suffices to prove the equality
\begin{equation}
\Delta_{i+1}^{} = \partial_+ \partial_- \Delta_i^{} - \partial_+
\Delta_i^{} \Delta_i^{-1} \partial_- \Delta_i^{} + \Delta_i^{}
\Delta_{i-1}^{-1} \Delta_i^{}. \label{b.11}
\end{equation}

Let $\rho = (\rho_{rs})_{r,s = 1, 2, \ldots}$ be the infinite
block matrix with the matrix elements defined by
\[
\rho_{rs} = \partial^{r-1}_+ \partial^{s-1}_- \beta,
\]
and let $\r{i}$ be a submatrix of $\rho$ formed by the matrix
elements $\rho_{rs}$ with $r, s \le i$. Recalling the definition of
$\Delta_{i}$ one can write $\Delta_{i} = |\r{i}|_{ii}$. Represent
the matrix $\r{i}$ in the form 
\begin{equation}
\r{i} = \left( \begin{array}{cc}
\r{i-1} & \t{i-1} \\
\s{i-1} & \rho_{ii}
\end{array} \right), \label{b.1}
\end{equation}
where the matrix elements of $1 \by (i-1)$ matrix $\s{i-1}$ and
$(i-1) \by 1$ matrix $\t{i-1}$ are given by
\[
\left( \s{i-1} \right)\!{}_a = \left( \r{i} \right)\!{}_{ia} =
\rho_{ia}, \qquad \left( \t{i-1} \right)\!{}_a = \left( \r{i}
\right)\!{}_{ai} = \rho_{ai}. 
\]
{}From equality (\ref{a.18}) we obtain the following expression: 
\begin{equation}
\Delta_{i} = \rho_{ii} - \s{i-1} \r{i-1}^{-1} \t{i-1}.
\label{b.3}
\end{equation}
Using this expression and representation (\ref{b.1}) we come to the
relation 
\begin{equation}
\r{i}^{-1} = \left( \begin{array}{c|c}
\r{i-1}^{-1} - \r{i-1}^{-1} \t{i-1} \Delta_{i}^{-1}
\s{i-1} \r{i-1}^{-1} &
- \r{i-1}^{-1} \t{i-1} \Delta_{i}^{-1} \\ \hline
- \Delta_{i}^{-1} \s{i-1} \r{i-1}^{-1} & \Delta_{i}^{-1}
\end{array} \right). \label{b.2}
\end{equation}
For the quasideterminant $\Delta_{i+1}$ one has
\[
\Delta_{i+1} = \rho_{i+1,i+1} - \s{i} \r{i}^{-1} \t{i}.
\]
Note that there are valid the following relations:
\[
\partial_+ \rho_{rs} = \rho_{r+1,s}, \qquad \partial_-
\rho_{rs} = \rho_{r, s+1}.
\]
Therefore, one can represent $\s{i}$ and $\t{i}$ as
\[
\s{i} = \left( \begin{array}{cc}
\partial_+ \s{i-1} & \partial_+ \rho_{ii}
\end{array} \right), \qquad 
\t{i} = \left( \begin{array}{c}
\partial_- \t{i-1} \\
\partial_- \rho_{ii}
\end{array} \right).
\]
Using this representation and relation (\ref{b.2}) one easily obtains
\begin{eqnarray}
\Delta_{i+1} &=& \partial_+ \partial_- \rho_{ii} - \partial_+
\s{i-1} \r{i-1}^{-1} \partial_- \t{i-1} \nonumber \\
&-& \left( \partial_+ \rho_{ii} - \partial_+ \s{i-1}
\r{i-1}^{-1} \t{i-1} \right) \Delta_i^{-1} \left(\partial_- \rho_{ii}
- \s{i-1} \r{i-1}^{-1} \partial_- \t{i-1} \right). \label{b.8}
\end{eqnarray}
Differentiating (\ref{b.3}) over $z^-$ we come to the relation
\begin{eqnarray}
\partial_- \Delta_i = \partial_- \rho_{ii} &-& \s{i-1}
\r{i-1}^{-1} \partial_- \t{i-1} \nonumber \\
&-& \left( \partial_- \s{i-1} - \s{i-1} \r{i-1}^{-1} \partial_-
\r{i-1} \right) \r{i-1}^{-1} \t{i-1}. \label{b.4}
\end{eqnarray}
It is easy to get convinced that
\[
\left( \r{i-1}^{-1} \partial_- \r{i-1} \right)\!{}_{ab} =
\left\{ \begin{array}{cl} 
\delta_{a, b+1}, & \quad b \ne i - 1; \\
\displaystyle \sum_{c=1}^{i-1} \left( \r{i-1}^{-1} \right)\!{}_{ac}
\rho_{ci}, & \quad b = i - 1. 
\end{array} \right.
\]
This relation implies the equality
\[
\left( \partial_- \s{i-1} - \s{i-1} \r{i-1}^{-1} \partial_-
\r{i-1} \right)\!{}_a = \Delta_i \delta_{i-1, a}.
\]
Therefore, from (\ref{b.4}) it follows that
\begin{equation}
\partial_- \Delta_i = \partial_- \rho_{ii} - \s{i-1}
\r{i-1}^{-1} \partial_- \t{i-1} - \Delta_i \left( \r{i-1}^{-1}
\t{i-1} \right)\!{}_{i-1}. \label{b.5}
\end{equation}
Now differentiate (\ref{b.3}) over $z^+$ and use the relation
\begin{equation}
\left( \partial_+ \r{i-1} \r{i-1}^{-1} \right)\!{}_{ab} =
\left\{ \begin{array}{cl} 
\delta_{a+1, b}, & \quad a \ne i - 1; \\
\displaystyle \sum_{c=1}^{i-1} \rho_{ic} \left( \r{i-1}^{-1}
\right)\!{}_{cb}, & \quad a = i - 1. 
\end{array} \right. \label{b.7}
\end{equation}
This gives
\begin{equation}
\partial_+ \Delta_i = \partial_+ \rho_{ii} - \partial_+ \s{i-1}
\r{i-1}^{-1} \t{i-1} - \left( \s{i-1} \r{i-1}^{-1} \right)\!{}_{i-1}
\Delta_i. \label{b.6}
\end{equation}
Using relations (\ref{b.5}) and (\ref{b.6}) one obtains
\begin{eqnarray}
\lefteqn{\partial_+ \Delta_i^{} \Delta_i^{-1} \partial_- \Delta_i^{}}
\nonumber \\
&\hskip 3.em&= \left( \partial_+ \rho_{ii} - \partial_+ \s{i-1}
\r{i-1}^{-1} \t{i-1} \right) \Delta_i^{-1} \left(\partial_- \rho_{ii}
- \s{i-1} \r{i-1}^{-1} \partial_- \t{i-1} \right) \nonumber \\
&&- \left( \partial_+ \rho_{ii} - \partial_+ \s{i-1}
\r{i-1}^{-1} \t{i-1} \right) \left( \r{i-1}^{-1} \t{i-1}
\right)\!_{i-1} \nonumber \\
&&- \left( \s{i-1} \r{i-1}^{-1} \right)\!{}_{i-1} \left(\partial_- \rho_{ii}
- \s{i-1} \r{i-1}^{-1} \partial_- \t{i-1} \right) \nonumber \\
&&+ \left( \s{i-1} \r{i-1}^{-1} \right)\!{}_{i-1} \Delta_i \left(
\r{i-1}^{-1} \t{i-1} \right)\!_{i-1}. \label{b.9}
\end{eqnarray}
The differentiation of (\ref{b.5}) over $z^+$ with account of
(\ref{b.6}) and (\ref{b.7}) gives
\begin{eqnarray}
\lefteqn{\partial_+ \partial_- \Delta_i^{} = \partial_+ \partial_- \rho_{ii}
- \partial_+ \s{i-1} \r{i-1}^{-1} \partial_- \t{i-1}} \nonumber \\
&\hskip 3.em& - \left( \partial_+ \rho_{ii} - \partial_+ \s{i-1}
\r{i-1}^{-1} \t{i-1} \right) \left( \r{i-1}^{-1} \t{i-1}
\right)\!_{i-1} \nonumber \\
&& - \left( \s{i-1} \r{i-1}^{-1} \right)\!{}_{i-1} \left(\partial_- \rho_{ii}
- \s{i-1} \r{i-1}^{-1} \partial_- \t{i-1} \right) \nonumber \\
&& + \left( \s{i-1} \r{i-1}^{-1} \right)\!{}_{i-1} \Delta_i \left(
\r{i-1}^{-1} \t{i-1} \right)\!_{i-1} - \Delta_i^{} \Delta_{i-1}^{-1}
\Delta_i^{}. \label{b.10}
\end{eqnarray}
Now with (\ref{b.8}), (\ref{b.10}) and (\ref{b.9}) we make sure that
relation (\ref{b.11}) is valid. Hence, relation (\ref{3.28}) is also
valid.

\end{document}